\documentclass[a4paper,11pt]{article}

\usepackage{fullpage}
\usepackage{lineno}
\usepackage[shortlabels]{enumitem}

\usepackage{amsmath, amssymb, amsthm}
\usepackage[mathic]{mathtools}
\usepackage{thm-restate}
\usepackage{graphicx,scalerel}

\usepackage[T1]{fontenc}

\usepackage{hyperref}
\hypersetup{
    colorlinks,
    linkcolor={red!50!black},
    citecolor={blue!50!black},
    urlcolor={blue!80!black}
}
\usepackage[capitalise]{cleveref}

\theoremstyle{definition}
\newtheorem{definition}{Definition}
\theoremstyle{plain}
\newtheorem{theorem}{Theorem}
\newtheorem{lemma}{Lemma}
\newtheorem{corollary}{Corollary}
\newtheorem{proposition}{Proposition}
\newtheorem{claim}{Claim}
\newtheorem{observation}{Observation}
\newtheorem*{maintheorem}{Main Theorem (informal)}

\crefname{observation}{Observation}{Observations}
\crefname{claim}{Claim}{Claims}

\makeatletter
\let\c@author\relax
\makeatother

\DeclareUnicodeCharacter{1EF3}{\`y}
\usepackage[backend=biber,isbn=false,url=false,doi=true,maxcitenames=3,maxbibnames=99,sortcites,style=numeric]{biblatex}
\DeclareFieldFormat{labelnumberwidth}{\mkbibbold{#1}}

\bibliography{strings-long,references}

\crefname{algocf}{Algorithm}{Algorithms}
\Crefname{algocf}{Algorithm}{Algorithms}

\usepackage{xspace}
\usepackage[dvipsnames]{xcolor}
\usepackage{orcidlink}
\usepackage{url}
\usepackage{soul}

\usepackage{tikz}
\usetikzlibrary{arrows, arrows.meta, calc, decorations.pathmorphing, decorations.pathreplacing, calligraphy, backgrounds, math, shapes, shapes.geometric, spy, positioning, quotes}

\makeatletter
\def\url@leostyle{%
  \@ifundefined{selectfont}{\def\UrlFont{\sf}}{\def\UrlFont{\small\ttfamily}}}
\makeatother
\DeclareMathOperator{\E}{\mathbb{E}}

\let\epsilon\varepsilon
\let\eps\varepsilon

\newcommand{\NN}{\ensuremath{\mathbb{N}}}

\newcommand{\FFF}{\mathcal{F}}

\newcommand{\GGG}{\mathcal{G}}
\newcommand{\HHH}{\mathcal{H}}

\usepackage{tikz}
\usetikzlibrary{decorations.pathreplacing}
\usetikzlibrary{plotmarks}
\usetikzlibrary{positioning,automata,arrows}
\usetikzlibrary{shapes.geometric}
\usetikzlibrary{decorations.markings}
\usetikzlibrary{positioning, shapes, arrows}
\definecolor{lightgray}{rgb}{0.83, 0.83, 0.83}

\usepackage[linesnumbered,ruled,vlined,boxed]{algorithm2e}
\newlength{\commentWidth}
\let\oldnl\nl
\newcommand{\nonl}{\renewcommand{\nl}{\let\nl\oldnl}}
\DontPrintSemicolon
\SetKwInOut{Input}{Input}\SetKwInOut{Output}{Output}

\newcommand{\A}{\mathcal A}
\newcommand{\G}{\mathcal G}

\newcommand{\F}{\mathcal F}

\newcommand{\bigO}{O} 
\newcommand{\prob}{\mathbb{P}}
\renewcommand{\Pr}{\prob}
\newcommand{\expect}{\mathbb{E}}

\newcommand{\w}{w}
\newcommand{\ie}{i.e.,\xspace}

\newcommand{\capped}[1]{\hstretch{2}{\hat{\hstretch{.5}{#1}}}}
\newcommand{\aas}{a.a.s.\@\xspace}

\DeclarePairedDelimiter{\abs}{\lvert}{\rvert}
\DeclareMathOperator*{\argmin}{arg\,min}
\DeclareMathOperator{\dd}{d\!} %
\DeclarePairedDelimiter\ceil{\lceil}{\rceil}

\newcommand{\pfrac}[2]{{\left(\frac{#1}{#2}\right)}}

\DeclareMathOperator{\ext}{ext}

\newcommand{\Erdos}{Erd\H{o}s}
\newcommand{\Renyi}{R\'enyi}
\newcommand{\ER}{\Erdos-\Renyi}

\tikzset{
	vertex/.style={
		circle,
		draw=black,
		solid,
		fill=black,
		minimum size=1.3mm,
		inner sep=0,
	},
	edge/.style={
		draw=black,
		thick,
	},
	border/.style = {
		draw=black,
		thin,
	},
}

\usepackage{todonotes}

\usepackage{etoolbox}

\nolinenumbers

\title{Giant Components in Random Temporal Graphs\footnote{An extended abstract of this work appeared in the Proceedings of the International Conference on Randomization and Computation (RANDOM 2023).}}

\author{
        Ruben Becker\thanks{Ca' Foscari University of Venice, Italy. Email: \texttt{rubensimon.becker@unive.it} \orcidlink{0000-0002-3495-3753}}
	\and
        Arnaud Casteigts\thanks{University of Geneva, Switzerland. The work was partially supported by the French ANR, project ANR-22-CE48-0001 (TEMPOGRAL). Email: \texttt{arnaud.casteigts@unige.ch} \orcidlink{0000-0002-7819-7013}}
	\and
        Pierluigi Crescenzi\thanks{Gran Sasso Science Institute, L'Aquila, Italy. Email: \texttt{pierluigi.crescenzi@gssi.it} \orcidlink{0000-0001-8789-3195}}
	\and
        Bojana Kodric\thanks{Ca' Foscari University of Venice, Italy. Email: \texttt{bojana.kodric@unive.it} \orcidlink{0000-0001-7242-0096}}
	\and
        Malte Renken\thanks{Technical University of Berlin, Germany. The work was partially supported by the German Research Foundation (DFG) under the project MATE (NI 369/17). Email: \texttt{m.renken@tu-berlin.de} \orcidlink{0000-0002-1450-1901}}
	\and
        Michael Raskin\thanks{LaBRI, CNRS, University of Bordeaux, France. The work was partially supported by the European Research Council (ERC) under the European Union’s Horizon 2020 research and innovation programme under grant agreement No~787367 (PaVeS). Email: \texttt{mraskin@u-bordeaux.fr} \orcidlink{0000-0002-6660-5673}}
	\and
        Viktor Zamaraev\thanks{University of Liverpool, United Kingdom. Email: \texttt{viktor.zamaraev@liverpool.ac.uk} \orcidlink{0000-0001-5755-4141}}
}

\begin{document}

\maketitle

\begin{abstract}
	
	A temporal graph is a graph whose edges appear only at certain points in time. 
	Recently, the second and the last three authors proposed a natural temporal analog of the \ER{} random graph model. The proposed model is obtained by randomly permuting the edges of an \ER{} random graph and interpreting this permutation as an ordering of presence times. It was shown that the connectivity threshold in the \ER{} model fans out into multiple phase transitions for several distinct notions of reachability in the temporal setting.
	
	In the present paper, we identify a sharp threshold for the emergence of a giant temporally connected component. We show that at $p = \log n/n$ the size of the largest temporally connected component increases from $o(n)$ to~$n-o(n)$. This threshold holds for both \emph{open} and \emph{closed} connected components, \ie{} components that allow, respectively forbid, their connecting paths to use external nodes.

\end{abstract}

{
\textbf{Keywords: }
{random temporal graph, Erd{\H{o}}s–R{\'e}nyi random graph, sharp threshold, temporal connectivity, temporal connected component}
}

\section{Introduction}

Many real-world networks vary with time,
as exemplified by the dynamic nature of today's social media, telecommunication, transportation, and interaction in general in a complex network.
Indeed, the examination of specific applications illustrates how networks endowed with temporal information enable more accurate and effective analysis of real-world systems compared to static networks \cite{tang2013applications}.

This insight has motivated plethora of studies focusing on network mod\-eling approaches that incorporate the time dimension \cite{KKK02,holme2012temporal, holme2019temporal}.
A widely used model for these networks is given by \emph{temporal graphs} (sometimes also called \emph{time-varying graphs}, \emph{evolving graphs}, or other names).
A temporal graph is a pair $\G = (G, \lambda)$, where $G=(V,E)$ is an \emph{underlying} (static) graph, and $\lambda$ is an \emph{edge labeling function} that assigns to every edge~$e \in E$ a set of time labels $\lambda(e) \subseteq \NN$ indicating when this edge is present.
This definition, although simple, already captures two important aspects that determine 
temporal networks. Namely, (a) the topology of the network defined by the underlying graph $G$; and (b) the schedule of edge availabilities represented by the labeling function $\lambda$.

Even though this model has gained much traction recently, the available tools for analyzing temporal graphs are still nowhere near the level of tools that have been developed for understanding static networks.
One of the main challenges is the fundamentally changed notion of \emph{reachability}.
In temporal graphs, reachability is naturally based on paths that traverse edges in
ascending time, a.k.a.\ \emph{temporal paths}. 
A first difference with standard paths is that temporal paths are inherently directed, regardless
of whether the graph itself is directed, due to the arrow of time.
Even more significantly, temporal reachability is
\emph{not transitive}, i.e., the fact that node $u$ can reach node~$v$
and node~$v$ can reach node~$w$ does not imply that~$u$~can reach~$w$.
The resulting non-composability is a source of complication
for structural studies, as well as a frequent source of computational hardness.
In fact,
many problems related to reachability are hard in temporal graphs,
even when their classical analogs are polynomial time solvable
--- see, for instance, the seminal paper by \textcite{KKK02}
on $k$-disjoint temporal paths (and many further examples appearing in
more recent works~\cite{Akr+17, CCMP20, waiting-time, minimization,
  separators, diameter, EMMZ21}).
As observed by~\textcite{BF03}, the fact that \emph{(temporally) connected components} do not form equivalence classes and intersect in non trivial ways implies, among other consequences, that finding one of maximum size is NP-hard.

\paragraph*{Random Models of Temporal Graphs.}
One of the most important tools in (static) network theory are random network models~\cite{newman2003structure}.
They allow reproducing characteristics of real networks and studying their statistical properties.
The random perspective enables prediction of properties, anomaly detection, identification of phase transitions, and other conclusions about the nature of typical networks. 

The cornerstone of random network theory is the \ER{} random graph model~\cite{NetworkScience}.
It has proven tremen\-dously useful as a source of insight into the structure of networks \cite{newman2018networks}.
An \ER{} random graph $G_{n,p}$ is obtained by placing an edge between each distinct pair
of~$n$~vertices%
\footnote{We use the terms \emph{vertex} and \emph{node} interchangeably.}
independently with probability $p$.
The study of this model was sparked by a series of seminal papers published by Erd\H{o}s and R\'{e}nyi starting in~1959~\cite{erdHos1959random, ER60-evolution, erdHos1961strength, erdHos1966existence}.
Since then, an important number of articles and books have been devoted to this model.
These results laid a solid foundation for the development of other models of more practical interest, including
the configuration model \cite{molloy1995critical, molloy1998size, newman2001random},
the small-world model \cite{watts1998collective}, and
the preferential attachment model \cite{barabasi1999emergence}.

The number of models of random \emph{temporal} networks proposed in the literature is still limited and
no systematic foundations are available \cite{holme2012temporal}.
In establishing such foundations, a natural question is: 
\emph{What is the temporal analog of the \ER{} random graph model?}
The answer to this question is not unique, as the time dimension can be incorporated 
in different ways~\cite{newman2018networks}.
Some candidates considered in the literature consider a sequence of independent \ER{} graphs, some others incorporate some dependencies in such a sequence (see for example ~\cite{chaintreau2007diameter,clementi2009broadcasting,clementi2010flooding, grindrod2010evolving, baumann2011parsimonious, zhang2017random, akrida2020fast}).

\paragraph*{Temporal \ER{} Random Graphs.}
Recently, another natural and more direct temporal analog of the \ER{} random graphs was proposed in \cite{CasteigtsRRZ21}. 
In this model, which we refer to as the \emph{temporal \ER{} random graph model}, a random temporal graph is obtained from an \ER{} random graph $G_{n,p}$ by assigning to each edge a unique label (presence time) according to a uniformly random permutation of its edges.
The main motivation is to obtain a temporal graph model whose properties (such as threshold values) can be directly compared to the classical \ER{} model, thereby highlighting the qualitative impact of the time dimension.
A systematic study of this model may also set a benchmark for practical models.

As already remarked, the time dimension leads to a number of distinctions between static and temporal graphs. Many of them come from the conceptual difference between the notions of path and temporal paths.
The reachability of a temporal graph is not symmetric (even in the undirected case) and not transitive, which is in stark contrast with static graphs.
Indeed, the results of \cite{CasteigtsRRZ21} revealed that even the notion of connectivity translates to a rich spectrum of phase transitions in the temporal setting.
Namely, at $p=\log n/n$, any \emph{fixed} pair of
vertices can asymptotically almost surely (\aas) reach each other; at $2\log n/n$, at least one
vertex (and in fact, any fixed vertex) can \aas reach all the others;
and at $3\log n/n$, \aas all the vertices can reach all others, \ie
the graph is temporally connected.

\paragraph*{Connected Components in Temporal \ER{} Random Graphs.}
Perhaps the most investigated aspects of \ER{} random graphs is the emergence of a ``giant'' connected component~\cite{Bollobas2001,frieze_karonski_2015}, which culminates in connectivity itself. The analogous question in a temporal setting is therefore natural.
Interestingly, the lack of transitivity makes the very definition of temporal components ambiguous. If the vertices of the component need temporal paths traveling \emph{outside} the component in order to reach each other, then the component is \emph{open}; otherwise, it is \emph{closed} \cite{BF03}.

Analyzing the emergence of (both types of) \emph{temporally} connected components in the above model presents technical challenges that cannot be overcome only by the tools developed in~\cite{CasteigtsRRZ21}.
These technical challenges and the importance of understanding connected components in temporal \ER{} random graphs motivated the present work.

\subsection{Contributions}
In this paper, we analyze 
the evolution of the largest connected component in a temporal \ER{} random graph with parameters $n$ and $p$, as $p$ increases (with $n \to \infty$).
Our main result is that, in contrast to static graphs, the phase transition occurs at~$p=\log n / n$.
At this point, the size of the largest component jumps from~$o(n)$ to~$n-o(n)$.

\begin{maintheorem}
	There exists a function $\eps(n) \in o(\log n / n)$ such that the size of a largest temporally connected component in a temporal \ER{} random graph is
	\begin{enumerate}
		\item[(i)] $o(n)$ \aas, if $p<\frac{\log n}{n} - \varepsilon(n)$; and
		\item[(ii)] $n-o(n)$ \aas, if $p>\frac{\log n}{n} + \varepsilon(n)$.
	\end{enumerate}
\end{maintheorem}

\noindent
Notably, the same threshold holds for both open and closed connected components, although showing the latter requires more effort.
We achieve these results by developing new techniques and combining them with strengthened versions of the tools from \cite{CasteigtsRRZ21}.
Informally, the new tools enable us to effectively contain the dependencies that exist between different time slices.
Thus they facilitate building graph structures witnessing a desired property in \emph{multiple independent phases}.

\subsection{Significance of the Results \& Techniques}

\textbf{Results.} Our main result reveals a qualitative difference between the evolution of connected components in static random graphs and temporal random graphs. 
The emergence of a giant component in (static) \ER{} graphs follows a
well-known pattern of events~\cite{ER60-evolution}. Below a
critical probability $p_0=1/n$, almost all the components are trees,
and no component is larger than $\bigO(\log n)$. Then, at~$p_0$, a
single ``giant'' component of size $\Theta(n^{2/3})$ arises. Then, at
$p = c/n > 1/n$, this component contains a constant fraction~$1-x/c$
of all vertices (with $0 < x < 1$ being defined through $xe^{-x} = ce^{-c}$).
As soon as $p\in\omega(1/n)$, the component contains all but $o(n)$
vertices. 
The case of directed static graphs is similar. Namely, for $p= c/n < 1/n$, a.a.s.\ all strongly connected components have size less than $3c^{-2}\log n$, and when $p= c/n > 1/n$, the graph contains a strongly connected component of size approximately
$\left(1- x/c\right)^2n$ (with $x$~as above)~\cite{Karp1990,frieze_karonski_2015}, which implies that this component contains all but $o(n)$ vertices when $p\in\omega(1/n)$.

In the temporal setting, we show that the phase transition occurs at
$p=\log n / n$. Namely, all components are of size $o(n)$ before that threshold
and there is one component of size $n - o(n)$ afterwards. The fact that this transition occurs later in the
temporal setting is not surprising, as the thresholds for
\emph{connectivity} is already known to be significantly smaller in
the static setting than in the temporal setting; namely, connectivity
occurs at $p = \log n / n$ in the static case (for both directed and
undirected graphs) versus $p = 3\log n / n$ for temporal
connectivity~\cite{CasteigtsRRZ21}. However, while these thresholds for
connectivity are within a multiplicative constant of
each other, our results show that in the case of connected components the static and the temporal threshold are of distinct
asymptotic orders.

\vskip2ex
\noindent
\textbf{Techniques.} In the temporal \ER{} model, the unicity of presence times for the edges causes delicate dependencies between past and future events.
To contain these dependencies, we introduce a multiphase analysis that consists of splitting the time interval into several phases where these dependencies are decoupled.
We believe that many further temporal graph properties will require such a multiphase analysis and could benefit from the tools developed here.
In constrast, the techniques from~\cite{CasteigtsRRZ21} are well suited for analyzing single-phase processes, where temporal paths do not interact across different time intervals (e.g. through composition).

In particular, we study a generalization of the temporal \ER{} model 
where a fixed base graph $G = K_n$ is replaced by an arbitrary graph of high minimum degree.
This generalization provides the possibility to ``encapsulate'' all dependencies on events occurring in some fixed ``short'' time phase into the choice of base graph, effectively eliminating the need to deal with these dependencies individually.
Another useful tool that we establish is the evolution of temporal paths
emanating from any of a set of source vertices.
Finally, one of our main technical results (which is used in order to prove the sharp threshold for closed connected components) essentially shows that, in the very early regime, there is only a small number of poorly connected vertices, and that these can be removed without compromising the connectivity of the remaining temporal graph. 
To achieve this, we partition vertices into a sequence of subsets and remove 
vertices in a subset if they are poorly connected to the vertices in the next subset in the sequence; this allows us to do the 'cleaning' in a controlled way to avoid cascading effect 
where removing poorly connected vertices from the graph can cause further vertices to become poorly connected.
This technique is of independent interest for ``bootstrapping'' multiphase analysis strategies similar to the ones used in this paper.

Although our techniques handle specific dependencies of temporal \ER{} graphs, they remain general enough to be adaptable to models with less dependencies, such as models where several appearances of an edge is possible and these appearances follow an exponential distribution (Poisson process). The reasons for this are exactly the same as discussed in~\cite{CasteigtsRRZ21}. Note, however, that weaker tools could suffice for such models, as past and future appearances of an edge are independent.

\subsection{Organization}

In Section~\ref{sec:Preliminaries}, we provide all necessary definitions, and introduce the random temporal graph models used in the paper.
In Section~\ref{sec: foremost forest algorithm}, we present the algorithm for constructing a foremost forest.
We also state a core technical theorem (\cref{theorem:general target set}) concerned with reachability between two sets of nodes in a temporal graphs.
Using this theorem, we then prove in~\cref{sec:threshold} that at $p = \log n/n$ the size of the largest \emph{open} connected component jumps from~$o(n)$ to~$n-o(n)$.
This also serves as a stepping stone towards~\cref{sec:threshold-closed}, where we extend our technique to also apply to \emph{closed} connected components. The proof is slightly more involved than for open components, as it requires further subdivisions of the phases. However, we show that both variants undergo essentially the same phase transition.

\section{Preliminaries}
\label{sec:Preliminaries}

In this paper, $[k]$ denotes the set of integers $\{1, \ldots, k\}$, and $[a, b]$ denotes either the discrete interval from $a$ to $b$, or the continuous interval from $a$ to $b$, the distinction being clear from the context.
All graphs are simple, i.e., without loops or multiple edges.
For a graph $G$, we denote by $V(G)$ and $E(G)$ its vertex set and edge set respectively.
We denote by $\delta(G)$ and $\Delta(G)$ the minimum and the maximum vertex degree of $G$ respectively.
As usual, $K_n$ denotes the complete $n$-vertex graph.

\subsection{Temporal Graphs}
A \emph{temporal graph} is a pair $(G, \lambda)$, where $G=(V,E)$ is a \emph{static} graph and $\lambda$ is a function
that assigns to every edge $e \in E$ a finite set of numbers, interpreted as presence times. The graph $G$
is called the \emph{underlying graph} of the temporal graph and the elements of $\lambda(e)$ are called the \emph{time labels} of $e$.
We will denote temporal graphs by calligraphic letters, e.g., by $\GGG$. Instead of $(G, \lambda)$ we will sometimes use the notation
$(V,E, \lambda)$ to denote the same temporal graph. In most cases, time labels will be elements of the real unit interval $[0,1]$.
Furthermore, in this paper, we restrict our consideration only to \emph{simple} temporal graphs\footnote{We remark that all our results can be directly transferred to another, closely related model of non-simple temporal graphs; see Section 6.1.2 in \cite{CasteigtsRRZ21}.},
i.e., temporal graphs in which every edge $e \in E$ is only present at a single point in time, i.e., $|\lambda(e)| = 1$.
We sometimes write $V(\GGG)$ and $E(\GGG)$ for the node and edge set of a temporal graph $\GGG$ respectively.

A temporal graph $\HHH=(V_\HHH, E_\HHH, \lambda_\HHH)$ is a \emph{temporal subgraph} of a temporal graph $\GGG=(V_\GGG, E_\GGG, \lambda_\GGG)$, if $V_\HHH\subseteq V_\GGG$, $E_\HHH\subseteq E_\GGG$ and $\lambda_\HHH(e)=\lambda_\GGG(e)$ for all $e\in E_\HHH$.
For a vertex set $S \subseteq V(\G)$, we denote by $\G[S]$ a temporal subgraph of $\G$
induced by $S$.
We use $\GGG_{[a,b]}$ to denote the temporal subgraph of $\GGG$ with the same node set $V_\GGG$, the edge set $E':=\{e\in E_\GGG:\lambda_\GGG(e)\in [a,b]\}$, and the time labeling function $\lambda_\GGG|_{E'}$ which is the restriction of $\lambda_\GGG$ to $E'$.

A \emph{temporal $(u, v)$-path} in $\GGG=(V, E, \lambda)$ between two nodes $u,v\in V$ is a sequence $u=u_0,u_1,\ldots, u_\ell=v$ such that $e_i=\{u_{i-1}, u_i\}\in E$ for each $i\in [\ell]$,
and time labels are increasing, i.e., $\lambda(e_1)< \ldots < \lambda(e_\ell)$. We call $\lambda(e_\ell)$ the \emph{arrival time} of the path.
A temporal $(u,v)$-path is called \emph{foremost $(u, v)$-path} if it has the earliest arrival time among all temporal $(u,v)$-paths.
If there exists a temporal $(u,v)$-path, we say that $u$ can reach $v$ (every vertex reaches itself).
A set~$S \subseteq V$ is said to \emph{reach~$v$} if at least one of its elements reaches~$v$.
In that case, a \emph{foremost $(S, v)$-path} in $\GGG$ is a temporal $(u, v)$-path with earliest arrival time among all $u\in S$.

A vertex $u\in V$ is called \emph{temporal source} in $\GGG=(V, E, \lambda)$ if there exists a temporal $(u,v)$-path for each $v\in V$.
Similarly, a vertex $v\in V$ is called \emph{temporal sink} in $\GGG$ if there exists a temporal $(u,v)$-path for each $u\in V$.

A temporal graph $\GGG=(V, E, \lambda)$ is \emph{temporally connected} if all nodes are temporal sources. We note that this also implies that all nodes are temporal sinks.
An \emph{open temporally connected component} or simply \emph{connected component} in $\GGG$ is an inclusion-wise maximal set $Z\subseteq V$
of nodes such that for every ordered pair of vertices $u,v\in Z$, there exists a temporal $(u,v)$-path in $\GGG$. We stress that such a temporal $(u,v)$-path can contain nodes from $V\setminus Z$.
If for every ordered pair $u,v\in Z$, there exists a temporal $(u,v)$-path in $\GGG[Z]$, then $Z$ is called
\emph{closed connected component}.

\subsection{Random Temporal Graph Models}
The model of \emph{temporal \ER{} random graphs} was introduced in~\cite{CasteigtsRRZ21}\footnote{In \cite{CasteigtsRRZ21}, this model was called Random Simple Temporal Graphs (RSTGs)} as a natural temporal generalization of the classical \ER{} model $G_{n,p}$ of random graphs. An $n$-vertex temporal \ER{} random graph with the parameter $p \in [0,1]$ is obtained by first drawing a \emph{static} random \ER{} $G_{n,p}$ and then defining a temporal order on its
edges by ordering them according to a uniformly random permutation. An equivalent and technically more convenient way of defining the temporal order on the edges is to draw, for every edge $e$, independently and uniformly at random a time label
$\lambda(e)$ from the unit interval $[0,1]$. Since the event that two edges get the same time label happens with probability 0,
all edge orderings induced by such random time labels are equiprobable. Therefore, as long as the absolute values of time labels are irrelevant (which is the case for the questions studied in \cite{CasteigtsRRZ21} and in the present paper), the two models are indeed equivalent. This latter model is denoted as $\F_{n,p}$. A possible way of generating objects from $\F_{n,p}$ is to first draw a temporal graph $\G=(G, \lambda)$ from $\F_{n,1}$ (thus the underlying graph $G$ is complete), and to then
consider $\G' = (G',\lambda') = (G,\lambda)|_{[0,p]}$, i.e., the temporal graph obtained from $\G$ by removing edges
with time labels greater than $p$. Observe that $G' \sim G_{n,p}$ and each time label $\lambda(e)$ is uniformly distributed
on $[0,p]$. Hence, $\G'$ is distributed according to $\F_{n,p}$ up to multiplying all labels by a factor of $\frac{1}{p}$, which we can ignore as it neither changes the relative order of time labels nor the absolute values of time labels are of any importance to us.
For similar reasons, for any $0 \leq a \leq b \leq 1$, up to rescaling time labels, the temporal subgraph $\G|_{[a,b]}$ is distributed according to $\F_{n,q}$, where $q=b-a$.

In order to overcome some technical challenges caused by interdependence of different temporal subgraphs, we define and study a natural generalization of $\F_{n,p}$ that we describe next.
For an $n$-vertex graph $G$ and a real value $p\in [0,1]$, we denote by $\FFF_p(G)$ the following random temporal graph model. A random temporal graph $\GGG=(V, E, \lambda) \sim \FFF_p(G)$ is obtained by
(1) independently and uniformly sampling a time label $\lambda'(e)\in [0,1]$ for every $e\in E(G)$, and
(2) setting $V=V(G)$, $E:=\{e\in E(G):\lambda(e)\le p\}$ and $\lambda(e) = \lambda'(e)$ for every $e\in E$.
We call $G$ the \emph{base graph} of $\FFF_p(G)$.
We observe that the $\FFF_{n,p}$ model is obtained when choosing the base graph to be the complete $n$-vertex graph $K_n$.

In what follows we sometimes implicitly assume that $n=|V|$ is sufficiently large without restating this assumption. We note that some of our estimates hold only for rather large values of $n$. We did not attempt to reduce these bounds, but instead focused on achieving best possible readability.

As a warm-up, we give a simple upper bound on $p$ which guarantees that $\G \sim \F_p(G)$ is temporally connected \aas

\subsection{A Two-Hop Bound on Temporal Connectivity}\label{app:twohop}
In this section, for certain graphs $G$, we provide a simple upper bound on $p$ which guarantees that $\G \sim \F_p(G)$ is temporally connected \aas The bound will be useful later in the paper. Its proof is a straightforward generalization of the proof of a similar result for $\F_{n,p} = \F_p(K_n)$ from \cite{CasteigtsRRZ21}. Nevertheless, we present the proof as a warm up to more involved analysis of the new model $\F_p(G)$. Recall that $\delta(G)$ is the minimum degree in the base graph $G$ whose $m=|E(G)|$ edges are retained with probability $p$.

\begin{lemma}\label{lem: two hops}
	Let $\GGG=(V, E, \lambda)\sim\FFF_{p}(G)$ be such that $p= \alpha \sqrt{\log n/n}$ and $\delta(G)\ge \beta n$ for $\alpha >0$ and $\beta >1/2$. Then an arbitrary node in $\GGG$ is a temporal source with probability at least $1 - n^{-\frac{\alpha^2(2\beta -1)}{2} + 1}$.
\end{lemma}
\begin{proof}
	Let $u\in V$. For any node $w\in V\setminus \{u\}$, we define
	\[
		V_{u,w}:=\{v\in V\setminus\{u,w\}: \{u,v\}\in E(G) \text{ and } \{v,w\}\in E(G)\}
	\]
	and observe that $|V_{u,w}|\ge 2(\delta(G) -1) - (n-2)\ge (2\beta -1)n$. We now define the following events (1) $R_w$ is the event that $u$ can reach $w$, (2) for $v\in V_{u,w}$, $S_{v,w}$ is the event that $u$ can reach $w$ in exactly two hops via $v$. Then clearly, for each $v\in V_{u,w}$, $\Pr[S_{v,w}]=p^2/2$.
	We thus get
	\begin{align*}
		\Pr[\overline{R_w}]
		\le \Pr\Big[\bigcap_{v\in V_{u,w}} \overline{S_{v,w}}\Big]
		= \prod_{v\in V_{u,w}} \Pr [ \overline{S_{v,w}} ]
		\le \Big(1 - \frac{p^2}{2}\Big)^{(2\beta -1)n}
		\le \\ e^{- \frac{p^2}{2}(2\beta -1)n}
		= n^{- \frac{\alpha^2(2\beta -1)}{2}}
	\end{align*}
	using the definition of $p$ in the last step.
	It remains to use a union bound over all  $w\in V\setminus \{u\}$ to conclude that $u$ can reach all nodes with probability at least $1 - n^{- \frac{\alpha^2(2\beta -1)}{2} + 1}$.
\end{proof}

\begin{corollary}[Two-Hop Bound]\label{cor: two hop}
	Let $p=\log n/\sqrt{n}$ and assume that $G$ is such that $\delta(G)\ge n - (\log n)^a$ for some $a\in\NN$. Then, $\GGG\sim \FFF_p(G)$ is temporally connected with probability at least $1-n^{-\frac{\log n}{4}}$.
\end{corollary}
\begin{proof}
	We use \cref{lem: two hops} with $\alpha=\sqrt{\log n}$ and $\beta=4/5$. Note that $\delta(G)\ge n - (\log n)^a \ge 4n/5$ for large enough $n$. Hence, together with a union bound, we obtain that all nodes are temporal sources in $\GGG$ with probability at least
	$
		1 - n^{-\frac{1}{4} \log n}
	$.
\end{proof}

\section{The Foremost Forest Algorithm}
\label{sec: foremost forest algorithm}

The main aim of this section is to present an algorithm for constructing a foremost forest and to prove a property of this algorithm.

Foremost forests play a crucial role in the development of our main technical tool: for a fixed set of vertices $S$ and a given number $k$,
the estimation of the minimum value of $p$ such that the vertices in $S$ can reach $k$ vertices in $\G = (V, E, \lambda) \sim  \F_p(G)$ \aas

We obtain such an estimation by examining the evolution of a foremost forest for $S$ in $\G$ via analysis of the execution of the formost forest algorithm on random temporal graphs.
To elaborate on this approach, let us consider $v \in V \setminus S$. We would like to estimate the probability that $S$ reaches $v$ in $\G$.
For this, we follow an approach similar to the one used in \cite{CasteigtsRRZ21}.
Let $\G'\sim \F_{1}(G)$ and observe that
the probability that $S$ can reach $v$ in $\G$ is equal to the probability that the temporal subgraph $\G'_{[0, p]}$ contains a temporal $(u, v)$-path $P$ for some node $u\in S$. This again is equivalent to the arrival time of $P$ in $\G'$ being at most $p$.
Therefore, the estimation of the parameter $p$ for which some node from $S$ can reach $v$ can be reduced to the estimation of the minimum arrival time of a foremost temporal path from $S$ to $v$ in $\G'\sim \F_{1}(G)$. A \emph{foremost forest for $S$ in $\G$} is a minimal temporal subgraph that preserves foremost reachabilities from $S$ to all other vertices reachable from $S$ in $\G$.
We proceed with the necessary formal definitions.
\begin{definition}\label{def: increasing temporal forest}
	Let $\GGG=(V, E, \lambda)$ be a temporal graph and let $S\subseteq V$ be a set of vertices.
	The graph $\GGG_F=(V_F, E_F, \lambda_F)$ is an \emph{increasing temporal forest for $S$}, if
	\begin{enumerate}[(a)]
	\item $\GGG_F$ is a temporal subgraph of $\GGG$,
	\item the graph $F=(V_F, E_F)$ is a forest (i.e.\ acyclic graph) with $|S|$ components,
	\item for each $s \in S$ there is a connected component~$T_s$ of~$F$ such that $s$~reaches all vertices of~$T_s$ in~$\GGG_F$.
	\end{enumerate}
\end{definition}
We are now ready to define (partial) foremost forests.
\begin{definition}\label{def: foremost forest}
	Let $\GGG=(V, E, \lambda)$ be a temporal graph, let $S\subseteq V$ be a set of vertices and let $\GGG_F=(V_F, E_F, \lambda_F)$ be an increasing temporal forest for $S$.
	\begin{enumerate}
			\item Then $\GGG_F$ is a \emph{partial foremost forest for $S$}, if, for all $v\in V_F \setminus S$, the unique temporal $(S, v)$-path in $\GGG_F$ is a foremost $(S, v)$-path in $\GGG$.
			\item A partial foremost forest for $S$ is a \emph{foremost forest for $S$} if $V_F$ contains all vertices reachable from $S$ in $\GGG$, i.e., $V_F=\{v\in V: \exists (u, v)\text{-temporal path in $\G$ for some }u\in S\}$.
			\item A (partial) foremost forest for $\{v\}$ is a \emph{(partial) foremost tree for $v$}.
		\end{enumerate}
\end{definition}

\paragraph*{The Algorithm.}
Next, we present an algorithm that, given a temporal graph $\GGG=(V, E, \lambda)$ and a set of nodes $S\subseteq V$ constructs a foremost forest $\GGG_F$ for $S$. This algorithm is a straightforward generalization of the foremost tree algorithm from~\cite{CasteigtsRRZ21}, where the input set $S$ is assumed to be singleton.

The idea of the algorithm similar to Prim's algorithm for minimum spanning trees in static graphs: Starting from $\GGG_F=(V_F,E_F, \lambda_F)=(S, \emptyset, \emptyset)$, which is trivially a partial foremost forest for $S$, the algorithm iteratively adds one node and one edge to $V_F$ and $E_F, \lambda_F$, respectively, until $\GGG_F$ becomes a foremost forest for $S$. The main difference to Prim's algorithm is that, in every iteration, the next edge to be added is chosen as the edge of minimum time label among all edges that extend the current increasing temporal forest. For brevity, we introduce the following notation.
We write $\GGG_F \cup e$ for adding the edge $e=\{u, v\}$ to $\GGG_F$, i.e., the result is the temporal graph $(V_F\cup\{u, v\}, E_F\cup\{e\}, \lambda_F\cup \{(e, \lambda(e))\})$.
The set of edges that extend the current partial forest can then be defined as
\begin{align*}
	\ext(\GGG_F)
	:=\{e=\{u,v\}\in E:
                u\in V_F, v\in V\setminus V_F,\text{ and } \GGG_F\cup e\text{ is an increasing temporal} \\ \text{forest for }S
		\}.
\end{align*}
We are now ready to state the algorithm and prove its correctness.

\medskip

\begin{algorithm}[H]
	\caption{\textsc{Foremost Forest}}
	\Input{Simple temporal graph $\GGG=(V, E, \lambda)$; set of nodes $S\subseteq V$.}
	\Output{Foremost forest for $S$.}
	\smallskip
	$k = |S| - 1$,
	$\GGG^{k}_F=(S, \emptyset, \emptyset)$\;
	\While{$\ext(\GGG^k_F)\neq \emptyset$}{
		$k := k + 1$\;
		$e_k := \argmin \{ \lambda(e) \mid e \in \ext(\GGG^{k-1}_F)\}$\;
		$\GGG^k_F := \GGG^{k-1}_F \cup e_k$\;
	}
  \Return{$\GGG^k_F$}
  \label{alg: foremost forest}
\end{algorithm}

\medskip

\subsection{Correctness of the Foremost Forest Algorithm}
\label{app:foremostTree}

The next lemma shows that \cref{alg: foremost forest} is correct and the time labels of the edges added to the forest
are increasing over the iterations. The proof of the lemma is very similar to the proof of the corresponding lemma for the foremost tree algorithm
from \cite{CasteigtsRRZ21}: for the sake of completeness, we fully present it here.

\begin{restatable}{lemma}{foremostTree}{\normalfont (Foremost Forest Algorithm Correctness)}\textbf{.}
  \label{lem:foremostTree}
  Let $\GGG=(V, E, \lambda)$ be a simple temporal graph and let $S\subseteq V$ be a set of nodes. Let $k$ be the final value of $k$ in Algorithm~\ref{alg: foremost forest} and let $\GGG^k_F$ be the temporal graph output by the algorithm. Then (i) $\GGG^k_F$ is a foremost forest for $S$ in $\GGG$ and (ii) $\lambda(e_{|S|}) \leq \ldots \leq \lambda(e_{k})$.
\end{restatable}

\begin{proof}
  In order to prove (i) we first observe that clearly $\GGG_F^k$ is an increasing temporal forest for $S$. Thus, in order to show that $\GGG_F^k$ is a foremost forest for $S$, it remains to prove that,
  \begin{enumerate}[(a)]
    \item $\GGG_F^k$ is a partial foremost forest, i.e., for all $v\in V(\GGG_F^k) \setminus S$, the unique temporal $(S, v)$-path in $\GGG_F^k$ is a foremost $(S, v)$-path in $\GGG$
    \item $\GGG_F^k$ is a foremost forest, i.e., $V(\GGG_F^k)=\{v\in V: \exists (u, v)\text{-temporal path in $G$ for some }u\in S\}$.
  \end{enumerate}
  We start by showing (a) by induction, i.e., we show for $\ell\in [|S|-1, k]$ that, for all $v\in V(\GGG_F^\ell) \setminus S$, the unique temporal $(S, v)$-path in $\GGG_F^\ell$ is a foremost $(S, v)$-path in $\GGG$.
  The statement is obvious for $\ell=|S|-1$. Now, let $|S| \leq \ell \leq k$ and assume the statement holds for $\ell-1$.
  Let $e_\ell = \{a,b\}$ be the edge added to $\GGG^{\ell - 1}_F$ to form $\GGG^{\ell}_F$, where $a \in V(\GGG^{\ell - 1}_F)$ and $b \in V\setminus V(\GGG^{\ell - 1}_F)$.
  As $b$ is the unique node in $V(\GGG^{\ell}_F)\setminus V(\GGG^{\ell - 1}_F)$, we only need to show that the temporal
  $(S,b)$-path in $\GGG^{\ell}_F$ is a foremost $(S,b)$-path in $\GGG$.
  Suppose that it is not, and let $P$ be a foremost $(S,b)$-path in $\GGG$. Let $e'$ be the first edge of $P$ with one endpoint in $V(\GGG^{\ell - 1}_F)$ and the other endpoint not in $V(\GGG^{\ell - 1}_F)$.
  Clearly, $\GGG^{\ell - 1}_F \cup e'$ is an increasing temporal forest for $S$ and thus $e'\in \ext(\GGG^{\ell - 1}_F)$. Furthermore, since $P$ is a foremost $(S, b)$-path in $\GGG$ and the temporal $(S, b)$-path in $\GGG_F^{\ell}$ is not, the arrival time of $P$ and thus $\lambda(e')$ is strictly less than $\lambda(e_\ell)$, contradicting the minimal choice of $e_\ell$.
  In order to prove (b), first note that clearly $V(\GGG_F^k)\subseteq\{v\in V: \exists (u, v)\text{-temporal path in $G$ for some }u\in S\}$. Now, assume for contradiction that there exists a node $v\in V\setminus V(\GGG_F^k)$ such that there exists a $(u, v)$-temporal path $P$ in $\GGG$ for some $u\in S$. Similar to before let $e=\{a,b\}$ be the first edge of $P$ with one endpoint in $V(\GGG^{k}_F)$ and the other endpoint not in $V(\GGG^{k}_F)$.
  Then, clearly, $e\in \ext(\GGG_F^k)$, contradicting termination of the algorithm.

  In order to prove (ii), assume that it does not hold and let $\ell \geq 2$ be the minimum index such that
  $\lambda(e_\ell) < \lambda(e_{\ell-1})$. Again, let $e_\ell = \{a,b\}$ and assume that $a \in V(\GGG_F^{\ell - 1})$ and $b \in V\setminus V(\GGG_F^{\ell - 1})$, and  let $e_i$ be the last edge of the $(S,a)$-path in $\GGG_F^{\ell - 1}$.
  Clearly $\lambda(e_i) < \lambda(e_\ell)$ because $\GGG_F^{\ell}$ is an increasing temporal forest. Hence, $e_i \neq e_{\ell - 1}$ and we obtain that both $e_\ell$ and $e_{\ell -1}$ are in $\ext(\GGG_F^{\ell - 2})$, a contradiction to the minimum choice of $e_{\ell -1}$.
\end{proof}

\subsection{Foremost Forest Target Set Reachability}
\label{app:general target set proof}

One of our main technical results is the following theorem which, for two given sets of nodes $S$ and $T$, quantifies the probability that a foremost forest grown from set $S$ reaches $T$.

\begin{theorem}[Foremost Forest Target Set Reachability]\label{theorem:general target set}
  Let
  \begin{itemize}
    \item $G$ be a graph of minimum degree $\delta(G)\ge n - (\log n)^a$ for some $a\in\NN$,
    \item let $S$ and $T$ be two sets of nodes in $G$ of cardinalities $s\in [(\log n)^{13}, n/2]$ and $t$, respectively,
    \item let $z=z(n)$ be a function with $\varepsilon\le z(n)\le 1 - \varepsilon$ for some constant $\varepsilon\in (0, 1)$, and
    \item let $\GGG\sim \FFF_{p}(G)$ with $p\ge \frac{z\log n -\log s}{n} +\frac{3\log\log n}{n}$.
  \end{itemize}
  Then the foremost forest algorithm from $S$ on $\GGG$ reaches
  $T$ with probability at least $1-\frac{5}{2}n^{-\log\log n} - e^{-\frac{t}{2n}(n^z-s)}$.
\end{theorem}

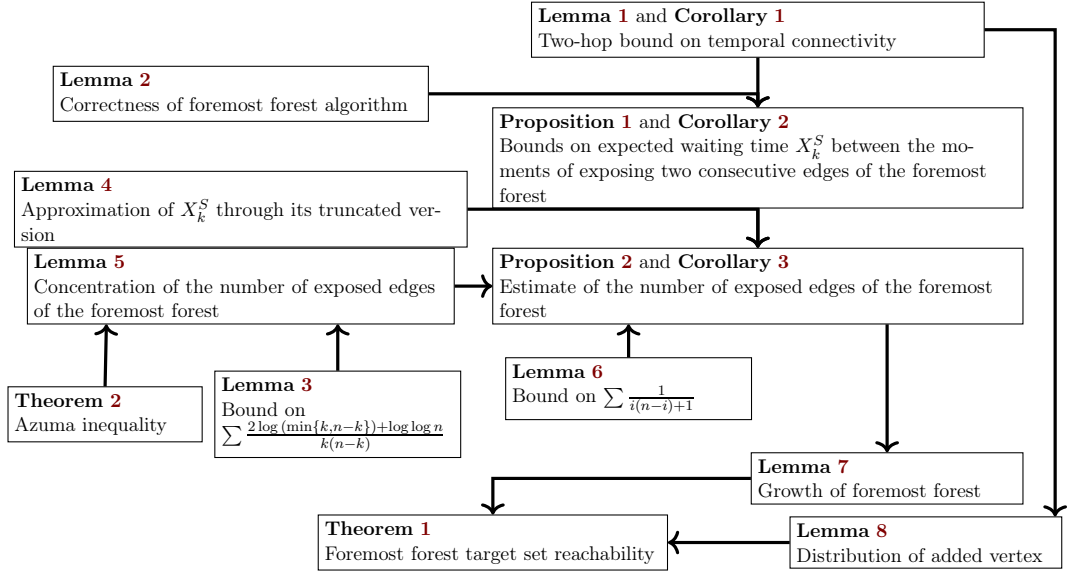
\begin{figure}[t]
    \centering
\tikzstyle{result} = [rectangle,draw,minimum height=1cm,minimum width=2cm]
\begin{tikzpicture}[x=0.25cm,y=0.25cm,scale=0.68, every node/.style={scale=0.68}]
\node[result,text width=7cm] (L3-3) at (0,5) {\textbf{\cref{lem:foremostTree}}\\Correctness of foremost forest algorithm};

\node[result,text width=8.5cm] (C2-2) at (40,10) {\textbf{\cref{lem: two hops}} and \textbf{\cref{cor: two hop}}\\Two-hop bound on temporal connectivity};

\node[result,text width=10cm] (P5-3) at (40,0) {\textbf{\cref{lem:ExpOfXk}} and \textbf{\cref{cor:ExpOfcappedXk}}\\Bounds on expected waiting time $X^{S}_{k}$ between the moments of exposing two consecutive edges of the foremost forest};

\node[result,text width=8.5cm] (L5-2) at (0,-4) {\textbf{\cref{lm:capped_eq_uncapped}}\\Approximation of $X^{S}_{k}$ through its truncated version};

\node[result,text width=10cm] (P5-8) at (40,-10) {\textbf{\cref{th:YConc}} and \textbf{\cref{th:YConcEasy}}\\Estimate of the number of exposed edges of the foremost forest};

\node[result,text width=8cm] (L5-6) at (0,-10) {\textbf{\cref{lm:YCapConcentrate}}\\Concentration of the number of exposed edges of the foremost forest};

\node[result,text width=3.5cm] (T5-5) at (-10.5,-20) {\textbf{\cref{the:Azuma}}\\Azuma inequality};

\node[result,text width=4.5cm] (L5-1) at (7.5,-20) {\textbf{\cref{lm:ck_bound}}\\Bound on \\ $\sum\frac{2 \log{(\min\{ k, n-k\})} + \log \log{n}}{k(n-k)}$};

\node[result,text width=4.5cm] (L5-7) at (30,-18) {\textbf{\cref{cl: favsum}}\\Bound on $\sum \frac{1}{i(n-i) + 1}$};

\node[result,text width=5cm] (L5-10) at (50,-25) {\textbf{\cref{lem: reachability}}\\Growth of foremost forest};

\node[result,text width=5cm] (L5-11) at (53,-30) {\textbf{\cref{lem: almost uniform}}\\Distribution of added vertex};

\node[result,text width=6.5cm] (T5-12) at (19.5,-30) {\textbf{\cref{theorem:general target set}}\\Foremost forest target set reachability};

\draw[->,very thick] (L3-3.east) -| (P5-3.north);
\draw[->,very thick] (C2-2.south) -- (P5-3.north);
\draw[->,very thick] (P5-3.south) -- (P5-8.north);
\draw[->,very thick] (L5-2.east) -| (P5-8.north);
\draw[->,very thick] (L5-6.east) -- (P5-8.west);
\draw[->,very thick] (T5-5.north) -- ([xshift=-2.6cm]L5-6.south);
\draw[->,very thick] (L5-1.north) -- ([xshift=1.875cm]L5-6.south);
\draw[->,very thick] (L5-7.north) -- ([xshift=-2.5cm]P5-8.south);
\draw[->,very thick] ([xshift=2.5cm]P5-8.south) -- (L5-10.north);
\draw[->,very thick] (C2-2.east) -| ([xshift=2.50cm]L5-11.north);
\draw[->,very thick] (L5-11.west) -- (T5-12.east);
\draw[->,very thick] (L5-10.west) -| (T5-12.north);
\end{tikzpicture}
    \caption{Overview of the proof of~\cref{theorem:general target set}.}
    \label{fig:reading_guide}
\end{figure}

The formal proof of \cref{theorem:general target set} is one of the technically more involved portions of this work.
It is divided into a number of lemmas and resides below in \cref{app:general target set proof};
for improved accessibility, a high level overview of the proof structure is depicted in \cref{fig:reading_guide}. 
The theorem is deduced from \cref{lem: reachability} and \cref{lem: almost uniform} that can be found in \cref{app:general target set proof}. \cref{lem: reachability} essentially constitutes a generalization of the foremost tree growth analysis from~\cite{CasteigtsRRZ21}, which estimates the number of vertices that a given vertex (referred to as a \emph{source}) reaches by specific time in $\F_{n,p}$. Besides the difference that in \cref{lem: reachability} we need to consider a fixed \emph{set of source} vertices, the main technical challenge here is that we have to consider the $\F_p(G)$ model rather than the basic $\FFF_{n,p}$ model, resulting in fewer edges per node. While \cref{lem: reachability} merely gives a statement over the number of nodes that are reached from a given source set, \cref{lem: almost uniform} gives the second crucial ingredient for proving \cref{theorem:general target set}.  It states that every new vertex reached by the foremost forest grown from $S$ (i.e., every new vertex added to the foremost forest) is distributed almost uniformly on the vertices that are not reached yet and this allows us to estimate the probability that the forest reaches the target set $T$.

\subsubsection{Proof of \cref{theorem:general target set}}
\label{sec: foremost forest analysis}
In this section, we analyze the Foremost Forest algorithm on a graph $\GGG\sim\FFF_1(G)$, where the base graph $G$ is such that $\delta(G)\ge n-(\log n)^a$ for some constant $a\in \NN$. In particular, the goal of the section is to prove~\cref{theorem:general target set} (see ~\cref{fig:reading_guide} for an overview of the proof). 
To this aim, we first want to estimate the time by which a typical foremost forest grown from a set $S$ of $s$ roots in $\GGG$ acquires a given number of vertices.
In this section, we globally assume that $s\ge (\log n)^b$ for some constant $b\ge a + 1$.
Sometimes we will need to assume stronger lower bounds on $b$, but this will then be explicitly stated.

We will consider the execution of the algorithm as a random process that reveals the edges of the resulting foremost forest one by one in the order in which they are added.
We restrict ourselves to running the algorithm until the set of sources $S$ reaches $n-s$ vertices in total.
In addition to the number of vertices that the foremost forest reaches by a given time, we will also analyze the distribution of the vertices that the forest acquires, i.e., we will show that every new vertex is distributed almost uniformly over the still unreached ones.

\paragraph*{Random Variables $X^S_k$ and $Y^S_k$.}
We use similar notation as in \cite{CasteigtsRRZ21} that is recapped in~\cref{sec: foremost forest algorithm}.
We denote the edges computed by the algorithm by $e^S_{s},\ldots, e^S_{n-s-1}$.
Note that for the purpose of more coherent notation, we start the numbering of the edges computed by the algorithm from $s$ (rather than from $1$).
We then define random variables
\[
  Y^S_{s-1} := 0 \text{ and }
  Y^S_k := \lambda(e^S_{k})\text{ for $k\in [s, n-s-1]$},
\]
where, extending the notation of \cref{alg: foremost forest} by a superscript denoting the input set $S$, we set $\GGG_F^{s-1, S} := (S, \emptyset, \emptyset)$
and, for every $k\in [s, n-s-1]$,
\begin{equation*}
  \begin{split}
    E^S_k   &:= V(\GGG_F^{k-1, S}) \times \big( V \setminus V(\GGG_F^{k-1, S}) \big)\\
    e^S_{k} &:= \arg\min \{ \lambda(e) ~|~ e \in E^S_k \text{ and } \GGG_F^{k-1, S} \cup \{e \} \text{ is an increasing temporal forest for }S \}, \\
    \GGG_F^{k, S}   &:= \GGG_F^{k-1, S} \cup \{ e^S_{k} \}.
  \end{split}
\end{equation*}
Note that by definition, for $k\in [s, n-s-1]$, $Y^S_k$ is the earliest time when the foremost forest for $S$ contains exactly $k-s+1$ edges, or equivalently the earliest time by which $S$ reaches $k+1$ vertices (vertices of $S$ included).
For $k\in [s, n-s-1]$, let $X^S_k$ be a random variable equal to $Y^S_k - Y^S_{k-1}$,
i.e., to the \emph{waiting time} between the edges $e^S_{k-1}$ and $e^S_k$. Clearly, we have
\[
  Y^S_k = \sum_{i=s}^k X^S_i
\]
for every $k \in [s, n-s-1]$.

We note that our assumption on the minimum degree of the base graph $G$ implies the following inequalities on the size of $|E^S_k|$, $k\in [s, n-s-1]$:
\[
    k \cdot (n - k)
    \ge |E^S_k|
    \ge k \cdot (n-k-\log^a n)
    = (1-\delta_k)\cdot k \cdot (n - k).
\]
where $\delta_k:=\frac{(\log n)^a}{n-k}$. Note that $\delta_k \leq \frac{(\log n)^a}{s+1} \leq \frac{(\log n)^a}{(\log n)^b+1} < 1$. We define $r_k\in [0,\delta_k]$ as the value for which $|E^S_k| = (1-r_k)\cdot k\cdot (n-k)$.

\paragraph*{Truncated Random Variables.}
We introduce the following \emph{truncated} versions of our main random variables $X^S_{s}, \ldots, X^S_{n-s-1}$ and $Y^S_{s}, \ldots, Y^S_{n-s-1}$,
which will be more convenient to bound later.
For $k\in [s, n-s-1]$, let
\[
  \capped{X}^S_k := \min \{ X^S_k, c_k \} \quad\text{ and }\quad
  \capped{Y}^S_k := \sum_{i=s}^{k} \capped{X}^S_i, 
\]\[
\quad \text{ where } \quad
  c_{k} := \frac{2 \log{(\min\{ k, n-k\})} + \log \log{n}}{k(n-k) \cdot s^{-1/3}}.
\]
We stress that the definition of $c_k$ (and thus $\capped{X}^S_k$ and $\capped{X}^S_k$) here extends the definition of $c_k$ from \cite{CasteigtsRRZ21} by the additional factor of $s^{1/3}$ (that equals 1 in the case of a single source).
The values of $c_k$ are chosen in such a way that on the one hand the sum of their squares is sufficiently small,
and on the other hand they are large enough to guarantee that the truncated variables coincide
with their original versions a.a.s. This is formalized in the following two lemmata.
\begin{lemma}[Bound on $\sum c_i$]\label{lm:ck_bound}
    It holds that
    \[
      \sum_{i=s}^{n-s-1} c_i^2 \leq \frac{64 (\log \log{n})^2(\log s)^2}{n^2s^{1/3}}.
    \]
\end{lemma}
\begin{proof}
  We get
  \begin{equation*}
    \begin{split}
      \sum_{i=s}^{n-s-1} c_i^2
      & = \sum_{i=s}^{n-s-1}\frac{\big(2 \log{(\min\{i, n- i\})} + \log \log{n}\big)^2}{(i\cdot (n- i))^2s^{-2/3}}
      \\ &\leq \sum_{i=s}^{n-s-1}\frac{\big(2 \log{(\min\{i, n- i\})} \cdot \log \log{n}\big)^2}{(i\cdot (n- i))^2s^{-2/3}}
      \\ &\leq \frac{4 (\log\log{n})^2}{s^{-2/3}} \Bigg( \sum_{i=s}^{\lfloor n/2 \rfloor}
      \frac{(\log i )^2}{ i^2\cdot (n/2)^2}
      + \sum_{i=\lfloor n/2 \rfloor+1}^{n-s-1} \frac{(\log(n- i))^2}{(n/2)^2(n- i)^2} \Bigg)
      \\ &\leq \frac{32 (\log \log{n})^2}{n^2s^{-2/3}} \sum_{i=s}^{\infty}\frac{(\log{i})^2}{i^2}
      \\ &\leq \frac{32 (\log \log{n})^2}{n^2s^{-2/3}} \int_{s-1}^{\infty}\frac{(\log{x})^2}{x^2}dx
      \\ &= \frac{32 (\log \log{n})^2}{n^2s^{-2/3}}\frac{(\log(s-1))^2 + 2 \log(s-1) + 2}{s - 1}
      \\ &\leq \frac{64 (\log \log{n})^2 (\log s)^2}{n^2s^{1/3}},
    \end{split}
  \end{equation*}
  assuming that $n$ and thus $s$ is sufficiently large.
\end{proof}

\begin{lemma}[$\hat X^S_k\approx X^S_k$]
  \label{lm:capped_eq_uncapped}
  With probability at least $1 - 1/(\log{n})^{s^{1/3}}$ the equality $\capped{X}^S_k=X^S_k$ holds for every $k\in [s, n-s-1]$.
\end{lemma}
\begin{proof}
  We observe that
  \begin{equation}\label{eq:prob_capped_uncapped}
  \begin{split}
    \prob \big[ \capped{X}^S_k \neq X^S_k \big]
    &= \prob \big[X^S_k > c_k \big]
    \leq (1 - c_k)^{\abs{E_k^S}}
    \leq (1 - c_k)^{k (n-k) \cdot(1-\delta_k)}
    \leq e^{-k (n-k) \cdot(1-\delta_k) \cdot c_{k}}
    \\&= \big(e^{-\log((\min\{ k, n-k\})^2 \log{n})}\big)^{s^{1/3}\cdot(1-\delta_k)}
    \le \big((\min\{ k, n-k\})^2 \log{n}\big)^{-s^{1/3}/2},
  \end{split}
  \end{equation}
  where we used that $\delta_k=(\log n)^a/(n-k)\le (\log n)^a/(s+1)\le (\log n)^{a-b}\le 1/2$, since $k\le n - s - 1$ and $b\ge a + 1$.
  Therefore
  \begin{align*}
      \sum_{k=s}^{n-s-1}
      \prob \big[\capped{X}^S_k \neq X^S_k \big] & \leq
      \sum_{k=s}^{n-s-1} \big((\min\{ k, n-k\})^2 \log{n}\big)^{-s^{1/3}/2}
      \\ &= \frac{1}{(\log{n})^{s^{1/3}/2}} \left( \sum_{k=s}^{\lfloor n/2 \rfloor} \frac{1}{k^{s^{1/3}}} +
      \sum_{k=\lfloor n/2 \rfloor+1}^{n-s-1} \frac{1}{(n- k)^{s^{1/3}}} \right)
      \\ &\leq \frac{2}{(\log{n})^{s^{1/3}/2}} \sum_{k=s}^{\infty}\frac{1}{k^{s^{1/3}}}
      \\ &\leq \frac{2}{(\log{n})^{s^{1/3}/2}} \int_{s-1}^{\infty}\frac{1}{x^{s^{1/3}}}dx
      \\ &= \frac{2}{(\log{n})^{s^{1/3}/2}} \frac{1}{(s^{1/3} - 1)\cdot (s-1)^{s^{1/3}-1}}
      \\ &\le \frac{2}{(\log{n})^{s^{1/3}/2}} \frac{1}{(s^{1/3} - 1)^{s^{1/3}}}
      \\ &\le \frac{1}{(\log{n})^{s^{1/3}/2} \cdot s^{s^{1/3}/4}},
  \end{align*}
  using that $n$ and thus $s$ is sufficiently large. It remains to use that $s\ge (\log n)^b$ and $b\ge a + 1 \ge 2$ to obtain the bound of $1/(\log n)^{s^{1/3}}$.
\end{proof}

\paragraph*{Expected Waiting Times.}
We will now estimate the expected time between exposing two consecutive edges of the foremost forest.
More specifically, we will bound the expected values of $X_k^S$ and $\capped{X}^S_k$
given the information revealed by the process in the first $k-1$ steps.
For every $k \in [s, n-s-1]$, let $\A^S_k$ be the $\sigma$-algebra generated
by the information revealed in the first $k$ iterations of the algorithm with sources $S$, i.e., by the knowledge of the first
$k$ edges $e^S_{s}, e^S_{s+1}, \ldots, e^S_{s + k - 1}$ and their time labels $Y^S_s, Y^S_{s+1}, \ldots, Y^S_{s+k-1}$.
Let also $\A^S_{s-1}$ be the trivial $\sigma$-algebra.
\begin{proposition}[{Bound on $\expect[X_k^{S}\mid\A_{k-1}^{S}]$}]\label{lem:ExpOfXk}
  For a set of vertices $S$ and every $k\in[s, n - s - 1]$ we have
  \begin{flalign*}
     \text{(i)}&& \frac{1-Y_{k-1}^{S}}{k(n-k) + 1}  &\leq  \expect[X_k^{S} \mid \A_{k-1}^{S}]  \leq  \frac{1}{k(n-k)(1-\delta_k) + 1};&
     \\ \text{(ii)}&&  \left( 1 - \frac{1}{(\log n)^{s^{1/3}}} \right) \cdot \frac{1-Y_{k-1}^{S}}{k(n-k) + 1}
     &\leq \expect[\capped{X}_k^{S} \mid \A_{k-1}^{S}]
     \leq \frac{1}{k(n-k)(1-\delta_k) + 1}.
  \end{flalign*}
\end{proposition}
\begin{proof}
  For every $k\in [s, n - s + 1]$ we define the function $\w_k^{S}: E_k^{S} \rightarrow [0,1]$ as
  \[
  \w^S_k(e) =
  \begin{cases}
    \lambda(e) - Y_{k-1}^{S}, & \quad \lambda(e) \geq Y_{k-1}^{S}\\
    \lambda(e) - Y_{k-1}^{S} + 1, & \quad \lambda(e) < Y_{k-1}^{S}.
  \end{cases}
  \]
  Notice that for any two edges $e,f \in E_k^{S}$ such that $\lambda(f) < Y_{k-1}^{S} \leq \lambda(e)$
  we have $\w_k^{S}(e) < \w_k^{S}(f)$. This together with the fact that the edge labels of $e_s^S,\ldots, e_{s+k+1}^S$ are increasing (see~\cref{lem:foremostTree}~(ii)) implies that $e_k^{S}$ is exactly
  the edge on which the minimum of $w_k^{S}$ is attained, that is,
  \begin{equation}\label{eq:ek_is_argmin}
    e_k^{S}
    = \arg\min \{\w_k^{S}(e) \mid e \in E_k^{S} \},
  \end{equation}
  and therefore, for every $k \in [s, n-s - 1]$,
  \begin{equation}\label{eq:Xk}
    X_k^{S}
    = \min\{w_k^{S}(e) \mid e \in E_k^{S}\}.
  \end{equation}
  Observe that upon exposure of edge $e_k^{S}$, we reveal some information about the time labels of the other edges in $E_k^{S}$.
  More precisely, we learn that these time labels are not contained in the interval $[Y_{k-1}^{S}, Y_k^{S}]$.
  Thus, if we inductively define the admissible range of $\lambda(e)$, $e \in E_k^{S}$, as
  \[
      I_k^{S}(e) :=
      \begin{cases}
        I_{k-1}^{S}(e) \setminus [Y_{k-2}^{S}, Y_{k-1}^{S}], & \quad e \in E_k^{S} \cap E_{k-1}^{S}\\
        [0,1], & \quad e \in E_k^{S} \setminus E_{k-1}^{S}
  \end{cases}
  \]
  then $\lambda(e)$ conditioned on $\A_{k-1}^{S}$ is uniformly distributed on $I_k^{S}(e)$.
  Let $\ell = \ell(e)$ be the unique index with $e \in E_\ell^{S} \setminus E_{\ell-1}^{S}$, i.e., the first iteration at which $e$ could have been added.
  Then we have $I_k^{S}(e) = [0,1] \setminus [Y_{\ell-1}^{S} , Y_{k-1}^{S}]$.
  It follows that $w_k^{S}(e)$ is uniformly distributed on its admissible range
  \[
    J_k^{S}(e) := \mathopen[ 0, Y_{\ell-1}^{S} - Y_{k-1}^{S} + 1 \mathclose]
  \]
  and clearly
    \begin{equation}\label{eq:inclusion}
    \mathopen[ 0, 1 - Y_{k-1}^{S} \mathclose] \subseteq J_k^{S}(e) \subseteq [0, 1] \,.
  \end{equation}

  Note that $X_k^{S}$ is a minimum of $k(n-k)(1-r_k)$ independent random variables $w_k^{S}(e), e \in E_k^{S}$,
  where for every edge $e \in E_k^{S}$ the value $w_k^{S}(e)$ is distributed uniformly on its own
  admissible range $J_k^{S}(e)$. Recall also that $r_k\in[0,\delta_k]$.
  Let $X_k'$ be the minimum of $k(n-k)(1-r_k)$ independent random variables distributed uniformly on
  $\left[0, 1-Y_{k-1}^{S} \right]$, and $X_k''$ be the minimum of $k(n-k)(1-r_k)$ independent
  random variables distributed uniformly on $[0, 1]$.
  Then, the inclusion \eqref{eq:inclusion} implies
  \[
    \prob[X_k' \geq t \mid Y_{k-1}^S] \leq \prob[X_k^{S} \geq t \mid A_{k-1}^{S}] \leq \prob[X_k'' \geq t],
    \]
  where $A_{k-1}^{S}$ is the event that specific
  edges $e^S_{s}, \ldots, e^S_{k-1}$ with time labels $Y^S_{s}, \ldots, Y^S_{k-1}$ are revealed in the
  first $k-s$ steps of the process. Therefore, by integrating and noting that
  the expected value of the minimum of $m$ independent variables distributed uniformly on $[0,a]$ is equal to $\frac{a}{m+1}$,
  we obtain
  \begin{equation}
    \begin{split}
      \frac{1-Y_{k-1}^{S}}{k(n-k) + 1}
      \leq
      \frac{1-Y_{k-1}^{S}}{k(n-k)(1-r_k) + 1}
      &=
      \expect[ X_k' \mid Y_{k-1}^S]
      \\&\leq
      \expect[ X_k^{S} \mid \A^v_{k-1}]
      \\&\leq
      \expect[ X_k'']
      =
      \frac{1}{k(n-k)(1-r_k) + 1}
      \\&
      \leq
      \frac{1}{k(n-k)(1-\delta_k) + 1}.
    \end{split}
  \end{equation}

  To prove the second part of the lemma we first note that by definition $\capped{X}_k^{S} \leq X_k^{S}$, and hence
  \[
    \expect[ \capped{X}_k^{S} \mid \A_{k-1}^{S} ]
    \leq
    \expect[ X_k^{S} \mid \A_{k-1}^{S} ]
    \leq
    \frac{1}{k(n-k)(1-\delta_k) + 1}.
  \]
    Therefore, it remains to show the lower bound on $\expect[ \capped{X}_k^{S} \mid \A_{k-1}^{S} ]$. For convenience, let us denote $M_k := k(n-k)(1-r_k)$.
  Then, we have
  \begin{equation*}
  \begin{split}
    \expect[ \capped{X}_k^{S} \mid \A_{k-1}^{S} ]
    &=
    \int_{0}^{\infty} \prob[\capped{X}_k^{S} \geq t \mid A_{k-1}^{S}] \dd t
    =
    \int_{0}^{c_k} \prob[\capped{X}_k^{S} \geq t \mid A_{k-1}^{S}] \dd t
    \\ &=
    \int_{0}^{c_k} \prob[X_k^{S} \geq t \mid A_{k-1}^{S}] \dd t
    \geq
    \int_{0}^{c_k} \prob[X_k' \geq t] \dd t
    \\ &=
    \int_{0}^{c_k} \left( 1 - \frac{t}{1 - Y_{k-1}^{S}} \right)^{M_k} \dd t
    \\ &=
    \frac{(1 - Y_{k-1}^{S}) + (c_k - (1 - Y_{k-1}^{S})) \left( 1 - \frac{c_k}{1 - Y_{k-1}^{S}} \right)^{M_k}}{M_k + 1}
    \\ & \geq
    \frac{(1 - Y_{k-1}^{S}) - (1 - Y_{k-1}^{S}) ( 1 - c_k )^{M_k}}{M_k + 1}
    \\ & =
    \frac{(1 - Y_{k-1}^{S}) \left( 1 - ( 1 - c_k )^{M_k} \right)}{M_k + 1},
  \end{split}
  \end{equation*}
    and the desired bound follows from the fact that
   $(1 - c_k)^{ k(n- k)(1-r_k)}  \leq (\log n)^{-s^{1/3}}$ following an analogous computation as in~\cref{eq:prob_capped_uncapped} and using that $\min\{k, n-k\}\ge s\ge \log n$.
\end{proof}

\begin{corollary}[{Final Bound on $\expect[X_k^{S}\mid\A_{k-1}^{S}]$}]\label{cor:ExpOfcappedXk}
  For a set of nodes $S$ of cardinality $s\ge (\log n)^b$ with $b\ge 3$, with probability at least $1-n^{-\log n /4}$, for every $k\in [s, n-s-1]$, it holds that
  \begin{flalign*}
    \text{(i)} &&   \left( 1 - \frac{\log{n}}{\sqrt{n}} \right) \cdot \frac{1}{k(n-k) + 1}  &\leq  \expect[X_k^{S} \mid \A_{k-1}^{S}]  \leq  \frac{1}{k(n-k)(1-\delta_k) + 1};&
    \\ \text{(ii)}  && \left( 1 - \frac{2\log n}{\sqrt{n}} \right) \cdot \frac{1}{k(n-k) + 1}  &\leq  \expect[\capped{X}_k^{S} \mid \A_{k-1}^{S}]  \leq  \frac{1}{k(n-k)(1-\delta_k) + 1}.
  \end{flalign*}
\end{corollary}
\begin{proof}
  Note first that according to~\cref{lem:ExpOfXk}, it remains to show the lower bounds.
  Recall that by \cref{cor: two hop}, $\GGG\sim \F_{\log n/\sqrt{n}}(G)$ is temporally connected with probability at least $1-n^{-\log n /4}$, in which case the bound $Y^S_k \leq Y^S_{n-s-1} \leq \log{n}/\sqrt{n}$ holds for every $S$ and $k\in [s, n-s-1]$. This together with~\cref{lem:ExpOfXk} implies the lower bound in (i).

  The lower bound in (ii) follows from the same bound on $Y^S_k$ and the fact that $(\log n)^{-s^{1/3}}\le 1/n^{\log \log n} \le \log n /\sqrt{n}$.
\end{proof}

\paragraph*{Deviation from Expectation.}
Next, we will bound the deviation of the truncated time when the foremost forest acquires $k$ edges from the expected value of accumulated truncated waiting times between the consecutive edges in the sequence of the first $k$ edges of the forest.
For this we require the following standard inequality by Azuma.
\begin{theorem}[Azuma's inequality~\cite{Azuma67}]\label{the:Azuma}
  Let $Z_0, Z_1, \ldots, Z_n$ be a martingale with respect to a filtration
     $\{\emptyset, \Omega\} = \A_0 \subset \A_1 \subset \ldots \subset \A_{n}$, and
      let $c_1, c_2, \ldots, c_n$ be non-negative numbers such that
      $$
        \sum_{i=1}^{n} \prob \left[ \abs{Z_i - Z_{i-1}} \geq c_i \right] = 0.
      $$
  Then
  \[
    \prob\bigl[ \abs{Z_n - Z_{0}} \geq \mu \bigr] \leq 2 \exp\left(\frac{-\mu^2}{2\sum_{i=1}^n c_i^2}\right).
  \]
\end{theorem}

\begin{lemma}[Concentration of $\hat{Y}^S_k$]
  \label{lm:YCapConcentrate}
  For a fixed set of vertices $S$ of cardinality $s\ge (\log n)^{b}$ with $b\ge 13$, with probability at least $1 - n^{-\log n}$ the inequality
  \[
    \left| \capped{Y}_k^{S} - \sum_{i=s}^k \expect \left[ \capped{X}_i^{S} \mid \A_{i-1}^{S} \right] \right| < \frac{\log\log n}{n}
  \]
  holds for all $k\in [s, n-s-1]$.
\end{lemma}

\begin{proof}
  Let us fix $k\in [s, n - s - 1]$
  and define a martingale $Z^S_{s-1}, Z^S_{s}, \ldots, Z^S_{k}$ with $Z^S_{s-1}:= 0$ and
  \[
    Z_t^{S} := \capped{Y}_t^{S} - \sum_{i=s}^t \expect [ \capped{X}_i^{S} \mid \A_{i-1}^{S} ]
         = \sum_{i=s}^t \capped{X}_i^{S} - \sum_{i=s}^t \expect[\capped{X}_i^{S} \mid \A_{i-1}^{S}],
  \]
  for $t \in [s, k]$.

  Since $0 \leq \capped{X}_i^{S} \leq c_i$, we have
  $0 \leq \expect[\capped{X}_i^{S} \mid \A_{i-1}^{S}] \leq c_i$,
  and therefore
  \[
  \prob\left[\abs*{Z_{i}^{S}-Z_{i-1}^{S}}>c_i \right]
  =
  \prob\left[ \abs*{\capped{X}_i^{S} - E[\capped{X}_i^{S} \mid\A_{i-1}^{S}]} >c_i\right]
  = 0
  \]
  holds for every $i\in [s, k]$.
  Furthermore, by \cref{lm:ck_bound}
  \[
    \sum_{i=s}^{k} c_i^2
    \leq
    \sum_{i=s}^{n-s-1} c_i^2
    \leq
    \frac{64 (\log \log{n})^2 (\log s)^2}{n^2s^{1/3}},
  \]
  and hence applying Azuma's inequality (\cref{the:Azuma}), we obtain that for sufficiently large $n$
  \begin{equation*}
    \begin{split}
      \prob \left[ |Z^S_k|\geq\frac{\log \log n}{n} \right]
      &\leq
      2\exp\left(
      \frac{-(\log \log n)^2}{n^2} \cdot
      \frac{1}{2\sum_{i=1}^{k}c_k^2}
      \right)
      \\ &\leq
      2\exp\left(
      \frac{-(\log \log n)^2}{n^2} \cdot
      \frac{n^2 s^{1/3}}{128 (\log \log{n})^2 (\log s)^2}
      \right)
      \\ &=
      2\exp\left(
      \frac{-s^{1/3}}{128 (\log s)^2}
      \right)
      \\ &\leq
      2\exp\left(\frac{-(\log n)^{b/3 - 2}}{128}\right)
      \\ &\leq
      n^{-\log n - 1},
    \end{split}
  \end{equation*}
  using that $b\ge 13$.
  The latter inequality together with the union bound over all $k\in [s, n-s-1]$ imply the desired result.
\end{proof}

\paragraph*{Deviation from $\sum_{i= s}^k \frac{1}{i(n-i) + 1}$.}
The following technical inequality will be useful in the rest of the section.
\begin{lemma}[Bound on $\sum \frac{1}{i(n-i) + 1}$]
\label{cl: favsum}
  For $k \in[s, n - 1]$, it holds that
  \begin{align*}
    &\frac{\log k - \log  s + \log (n -  s + 1) - \log(n - k)}{n} - \frac{3}{n}
    \\&\qquad \leq \sum_{i= s}^k \frac{1}{i(n-i) + 1} \leq \sum_{i= s}^k \frac{1}{i(n-i)}
    \\&\qquad \qquad\leq \frac{\log k - \log  s + \log (n -  s + 1) - \log(n - k)}{n} + \frac{3}{n}.
  \end{align*}
\end{lemma}
\begin{proof}
  We have that
  \begin{align*}
    \sum_{i= s}^k \frac{1}{i(n-i) + 1}
    &\leq \sum_{i=1}^k \frac{1}{i(n-i)}
    \\ &= \frac{1}{n} \sum_{i= s}^k \left( \frac{1}{i} + \frac{1}{n-i} \right)
    \\ &= \frac{1}{n} \left(\sum_{i=1}^{k} \frac{1}{i} - \sum_{i=1}^{ s - 1} \frac{1}{i} + \sum_{i=1}^{n -  s} \frac{1}{i} - \sum_{i=1}^{n-k-1} \frac{1}{i}\right)
    \\ &\leq \frac{\log k + 1 - \log  s + \log (n- s) + 1 - \log(n - k)}{n}
    \\ &\leq \frac{\log k - \log  s + \log (n -  s + 1) - \log(n - k)}{n} + \frac{3}{n}
    \intertext{as well as}
    \sum_{i= s}^k \frac{1}{i(n-i) + 1}
    &\geq  \sum_{i= s}^k \frac{1}{i(n+1-i)}
    \\&= \frac{1}{n+1}  \sum_{i= s}^k \left( \frac{1}{i} + \frac{1}{n+1-i} \right)
    \\ &= \frac{1}{n+1} \left( \sum_{i=1}^{k} \frac{1}{i} - \sum_{i=1}^{ s - 1} \frac{1}{i} + \sum_{i=1}^{n + 1 -  s} \frac{1}{i} - \sum_{i=1}^{n - k} \frac{1}{i} \right)
    \\ &\geq \frac{\log k - (\log  s + 1) + \log (n- s+1) - (\log(n-k) + 1)}{n+1}
    \\ &= \frac{\log k - \log  s + \log (n -  s + 1) - \log(n-k)}{n} \cdot\left(1 - \frac{1}{n+1}\right) - \frac{2}{n + 1}
    \\ &\geq \frac{\log k - \log  s + \log (n -  s + 1) - \log(n - k)}{n} - \frac{3}{n}.\qedhere
  \end{align*}
\end{proof}

\begin{proposition}[Deviation of $Y^S_k$ from $\sum \frac{1}{i(n-i) + 1}$]\label{th:YConc}
  For a set of vertices $S$ of cardinality $ s\ge (\log n)^b$, where $b\ge 13$,
  with probability at least $1 - 2n^{-\log \log n}$,
  for every $k\in [ s, n- s-1]$, we have
  \[
  \left| Y_k^{S} - \sum_{i=s}^k \frac{1}{i(n-i) + 1} \right|
  < \frac{2\log\log n}{n}.
  \]
\end{proposition}
\begin{proof}
  We will first prove that, with probability at least $1 - 2n^{-\log n/4}$, for a fixed set $S$ of $s$ vertices, for all $k\in [ s, n- s-1]$,
  \begin{align}\label{eq: trunc}
      \left| \capped{Y}_k^{S} - \sum_{i=s}^k \frac{1}{i(n-i) + 1} \right| < \frac{2\log\log n}{n}.
  \end{align}
  From there it will be straightforward to reach the statement of the proposition.

  We note that, for $i\in [s, k]$, we have that $\delta_i = \frac{(\log n)^a}{n - i} \le \frac{(\log n)^a}{s} \le \frac{1}{\log n}$ using that $b\ge a + 1$. Hence,
  $1/(1-\delta_i) \le  1 + 2/\log n$ and thus
  \[
    \frac{1}{i(n-i)(1-\delta_i) + 1}
    \le \Big(1 + \frac{2}{\log n}\Big) \cdot \frac{1}{i(n-i)+ 1}.
  \]
  By \cref{cor:ExpOfcappedXk}~(ii), with probability at least $1-n^{-\log n/4}$,
  for every set of vertices $S$ with $ s\ge (\log n)^{b}$, where $b\ge 3$ and every $k\in[s, n -  s - 1]$ we thus have
  \begin{equation}\label{eq:diffYhatExpHarmonic}
    \left|\sum_{i= s}^k \expect[\capped{X}^S_i \mid \A^S_{i-1}] - \sum_{i= s}^k \frac{1}{i(n-i) + 1} \right|
    \le \max
      \left\{
        \frac{2\log n}{\sqrt{n}}, \frac{2}{\log n}
      \right\}
    \cdot \sum_{i= s}^k \frac{1}{i(n-i) + 1}
  \end{equation}
  If $s\ge (\log n)^b$ with $b\ge 13$, we have by \cref{lm:YCapConcentrate} that with probability at least $1 - n^{-\log n}$,
  for every $k\in [ s, n- s-1]$,
  \begin{equation}\label{eq:Yhat}
      \left|\capped{Y}^S_k - \sum_{i= s}^k \expect[\capped{X}^S_i \mid \A^S_{i-1}] \right|
      \leq \frac{\log\log n}{n}.
  \end{equation}
  Hence, \eqref{eq:diffYhatExpHarmonic} and \eqref{eq:Yhat} together yield that, with probability at least $1 - 2n^{-\log n/4}$, for every $k\in [ s, n- s-1]$,
  \begin{align*}
          \left| \capped{Y}_k^{S} -  \sum_{i=s}^k \frac{1}{i(n-i) + 1} \right| 
          \hskip-2cm&
          \\&\le \left|\capped{Y}^S_k - \sum_{i= s}^k \expect[\capped{X}^S_i \mid \A^S_{i-1}] \right|
      + \left|\sum_{i= s}^k \expect[\capped{X}^S_i \mid \A^S_{i-1}] - \sum_{i= s}^k \frac{1}{i(n-i) + 1} \right|
    \\&\le
    \frac{\log\log n}{n} + \frac{2}{\log n} \cdot \sum_{i= s}^k \frac{1}{i(n-i) + 1}.
  \end{align*}
  Applying \cref{cl: favsum} and assuming that $n$ is sufficiently large enough, we get
  \begin{align*}
    \frac{2}{\log n} \cdot \sum_{i= s}^k \frac{1}{i(n-i) + 1}
    \leq
    \frac{2}{\log n} \cdot
    \left(
      \frac{2 \log n}{n} + \frac{3}{n}
    \right)
     \leq\frac{\log\log n}{n}
  \end{align*}
  and this completes the proof of~\eqref{eq: trunc}.

  Now by~\cref{lm:capped_eq_uncapped} and the assumption on $b$, and thus on $s$, it holds that the truncated and non-truncated random variables coincide with probability at least $1-n^{-\log \log n}$. Using that $2n^{-\log n/4}\le n^{-\log\log n}$ concludes the proof.
\end{proof}
The latter statement together with \cref{cl: favsum} immediately implies the following corollary.
\begin{corollary}[Final Deviation of $Y^S_k$]\label{th:YConcEasy}
  For a set of vertices $S$ of cardinality $ s\ge(\log n)^b$ with $b\ge 13$, with probability at least $1 - 2n^{-\log\log n}$, for every $k\in[ s, n- s-1]$ we have
  \[
  \abs*{Y_k^S - \frac{\log k - \log  s + \log (n -  s + 1) - \log(n - k)}{n}} < \frac{2\log\log n + 3}{n}.
  \]
\end{corollary}

\paragraph*{Time to Reach $n^z$ Nodes.}
Finally, we use \cref{th:YConcEasy} to prove the following lemma.
We will later reuse this also in the proof of \cref{thm:closed-component}.
\begin{lemma}[Foremost Forest Growth]\label{lem: reachability}
  Let
    \begin{itemize}
      \item $S$ be a set of nodes of cardinality $s\in [(\log n)^{b}, n/2]$ with $b\ge 13$,
      \item let $z=z(n)$ be a function with $\varepsilon\le z(n)\le 1 - \varepsilon$ for some constant $\varepsilon\in (0, 1)$ such that $d:=\ceil*{n^z} \leq n-s$,
      \item let $\GGG\sim \FFF_{p}(G)$ with $p\ge \frac{z\log n -\log s}{n} +\frac{3\log\log n}{n}$.
  \end{itemize}
    Then the foremost forest algorithm from $S$ on $\GGG$ reaches at least $d$ vertices with probability at least $1 - 2n^{-\log \log n}$.
\end{lemma}
\begin{proof}
  Clearly it suffices to consider the case $d > s$.
  To prove the lemma, it suffices to show that, with probability at least $1 - 2n^{-\log\log n}$, we have
  \[
      \left| Y^S_{d-1} - \frac{z\log n - \log s}{n} \right|
      <
      \frac{3\log \log n}{n}.
  \]
  Let
  \begin{align*}
    L \coloneqq{}& \log (d -1) - \log s + \log (n -  s + 1) - \log(n - d + 1).
  \end{align*}
  Then, for sufficiently large $n$,
  \begin{align*}
    L\le{}& \log (n -  s + 1) - \log s + \log\left( n^z \right) - \log\left(n-n^z\right)
    \\ \le{}& \log n - \log s + \log\left( n^z \right) - \log\left(n-n^z\right)
    \\ ={}& z \log n  - \log s - \log\left(1-n^{z-1}\right)
    \\ \leq{}& z \log n  - \log s + \log\left(1+\frac{n^{-\eps}}{1 - n^{-\eps}}\right)
    \\ \le{}& z \log n  - \log s + \frac{n^{-\eps}}{1 - n^{-\eps}}
    \\ \le{}& z \log n - \log  s + 1.
  \end{align*}
  Similarly, for sufficiently large $n$, and using that $s\le n/2$,
  \begin{align*}
    L\ge{}& \log\left(n^z - 1\right) - \log s + \log \left(\frac{n}{2}\right) - \log n
    \\ ={}& z \log n  - \log  s + \log\left(1-n^{-z}\right) - \log 2
    \\ \ge {}& z \log n  - \log  s + \log\left(1-n^{-\eps}\right)- \log 2
    \\ \ge {}& z \log n  - \log  s - 2\log 2
    \\ \ge{}& z \log n - \log  s - 2,
  \end{align*}
  where in the third step we used $z\ge \eps$.
  Then we conclude by \cref{th:YConcEasy} that with probability at least $1 - 2n^{-\log\log n}$
  \begin{align*}
    \abs*{ Y^S_{d-1} - \frac{{z\log n - \log s}}{n}}
    &\leq
    \abs*{ Y^S_{d-1}- \frac{L}{n} }
    +
    \abs*{ \frac{L}{n} - \frac{z\log n - \log s}{n} }
    \\ &\leq
    \frac{2\log\log n + 3}{n} + \frac{2}{n}
    \le
    \frac{3\log \log n}{n}
  \end{align*}
  holds.
\end{proof}

\paragraph*{Distribution of New Vertices.}
For $k\in [s, n-s -1]$, we call the node $u\in e_k^S\setminus \GGG_F^{k-1,S}$ the \emph{$k$-added vertex}, as it is added when the set $S$ without the newly added vertex has size $k$.
Recall that $\A_{k-1}^S$ refers to the knowledge of the edges of $\GGG_F^{k-1,S}$, including their time labels.
In the next lemma we show that the $k$-added vertex is distributed almost uniformly on $V\setminus V(\GGG_F^{k-1,S})$.
\begin{lemma}[$k$-Added Vertex Distribution]\label{lem: almost uniform}
  Let $ k \in [s, n-s-1]$ and $\gamma = \gamma(n) := 2\left(\frac{\log n}{\sqrt{n}} + \frac{(\log n)^a}{s}\right)$. 
  If~$Y_{n-s-1}^S \leq \frac{\log n}{\sqrt{n}}$,
  then it holds that
  \[
    \sup_{u \in V\setminus V(\GGG_F^{k-1,S})} \abs*{ \prob \left[ u \in e_k^S \mid \A_{k-1}^S \right] - \frac{1}{n-k}} \leq \frac{\gamma}{n-k}.
  \]
\end{lemma}
\begin{proof}
  Define $\gamma_1 := \frac{\log n}{\sqrt{n}}$.
  Recall that for each $k \in [s, n-s-1]$ we have
  \[
    e^S_k = \arg\min \{ \w^S_k(e) \mid e \in E^S_k \},
  \]
  where the variable $\w^S_k(e)$ is uniformly distributed on $J_k^S(e)$,
  which, by \eqref{eq:inclusion} and our assumption on $Y^S_{n-s-1}$, is an interval of the form $[0, w_e]$ with $1-\gamma_1 \le w_e \le 1$.
  Further recall that $|E^S_k|\in[(1-\delta_k)k (n-k),\, k(n-k)]$.
  
  Conditioned on $\A_{k-1}^S$, the probability~$p_e$ of the event~$e = e_S^v$ is then
  \begin{align*}
    p_e &= \int_{0}^{w_e} \frac{1}{w_e} \prod_{e' \in E^S_k \setminus \{e\}} \hspace{-2ex} \prob[\w^S_k(e') > x] \dd x
    = \int_{0}^{w_e} \frac{1}{w_e} \prod_{e' \in E^S_k \setminus \{e\}} \hspace{-2ex} \max\left\{0, \frac{w_{e'} - x}{w_{e'}}\right\} \dd x
    \\
    &\leq \int_0^1 \frac{1}{w_e} (1-x)^{\abs{E^S_k}-1} \dd x
    = \frac{1}{w_e \cdot \abs{E^S_k}} \leq \frac{1}{(1-\gamma_1) \cdot (1-\delta_k) \cdot k (n-k)}.
  \end{align*}
  On the other side, when setting $w_{\min} := \min\{w_e \mid e \in E^S_k\}$, then $p_e$~is lower-bounded by
  \begin{align*}
    p_e &\geq \int_0^{w_{\min}} \frac{1}{w_e} \left(\frac{w_{\min} - x}{w_{\min}}\right)^{\abs{E^S_k}-1} \dd x
    = \frac{w_{\min}}{w_e \cdot \abs{E^S_k}}
    \geq \frac{1-\gamma_1}{k (n-k)}.
  \end{align*}
  
  We now recall that each vertex $v$ has at most $(\log n)^a$ non-neighbors in the base graph $G$ and thus each vertex $u\in V\setminus V(\GGG_F^{k-1,S})$ has $n_u$ neighbors in $V(\GGG_F^{k-1,S})$, where
  \begin{align}\label{eq:nubound}
    k
    \ge n_u
    \ge k - (\log n)^a
    \ge k \cdot \Big(1 - \frac{(\log n)^a}{s}\Big)
    =: k \cdot (1-\gamma_3).
  \end{align}
  As
  $
    \prob[u\in e_k^S \mid \A_{k-1}^S]
    = \sum_{e\in E\cap (V(\GGG_F^{k-1,S}\times \{u\}))} p_e
  $,
  by combining our bounds on~$p_e$ with~\eqref{eq:nubound}, we obtain
  \[
    (1-\gamma_1)(1-\gamma_3)\cdot \frac{1}{n-k}
    \le \prob[u\in e_k^S \mid \A_{k-1}^S]
    \le \frac{1}{(1-\gamma_1)(1-\delta_k)}\cdot \frac{1}{n-k}.
  \]
  Notice now that
  \begin{align}\label{eq:uniformdown}
    (1-\gamma_1)(1-\gamma_3)\cdot \frac{1}{n-k}
    \ge (1 - \gamma_1 - \gamma_3)\cdot \frac{1}{n-k}
    \geq (1 - \gamma)\cdot \frac{1}{n-k}.
  \end{align}
  On the other hand, $\delta_k= \frac{(\log n)^a}{n - k}\le \frac{(\log n)^a}{s}=\gamma_3$ and thus
  \begin{align}\label{eq:uniformup}
    \begin{split}
      \frac{1}{(1-\gamma_1)(1-\delta_k)}\cdot \frac{1}{n-k} 
      &\le \frac{1}{1-\gamma_1-\gamma_3} \cdot \frac{1}{n-k}
      \\ &\le \left(1 + 2(\gamma_1+\gamma_3) \right) \cdot \frac{1}{n-k}
      \\ &= (1 + \gamma) \cdot \frac{1}{n-k}.
    \end{split}
  \end{align}
  Now, \eqref{eq:uniformdown} and \eqref{eq:uniformup} together show the claim.
\end{proof}

\paragraph*{Target Set Version of Foremost Forest Growth.}
Using the above observation that the $k$'th new vertex is almost uniformly distributed among the non-reached vertices, we are now ready to prove a result lower bounding the probability that the Foremost Forest algorithm started from a set $S$ hits a fixed set $T$ of large enough size. That is, we can finally prove~\cref{theorem:general target set}.

\begin{proof}
  Assume for now that $Y_{n-s-1}^S \leq \frac{\log n}{\sqrt{n}}$.
  For $k\in [s, n - s - 1]$, let $P_k$ denote the event that the $k$-added vertex does not belong to $T$ and let $Q_k:=\bigcap_{i=s}^k P_i$ denote the event that all new vertices up to the $k$-added one do not belong to $T$.
  By~\cref{lem: almost uniform}, the $k$-added vertex is distributed almost uniformly on the vertices not reached yet.
  Thus, for $i\in [s, k]$, the probability that the $i$-added vertex belongs to $T$ conditioned on the events that the previous ones did not is $\Pr[\overline{P_i} \mid Q_{i-1}]\ge \frac{t\cdot(1-\gamma(n))}{n - i}\ge \frac{t}{2n}$ using the facts that $n-i\le n$ and $\gamma=\gamma(n)\le 1/2$ for $n$ large enough (see~\cref{lem: almost uniform} for the definition of $\gamma$).
  Hence, we get
  \begin{align*}
    \Pr[Q_k]
    = \prod_{i=s}^k \Pr[P_i \mid Q_{i-1}]
    \le \left(1-\frac{t}{2n}\right)^{k - s}
  \end{align*}
  Let us now define $A$ as the event that
  $S$ can reach at least $d=\lceil n^z\rceil$ vertices. Using~\cref{lem: reachability} we obtain that $A$ occurs with probability at least $1-2n^{-\log \log n}$.
  Then, the probability that none of the new vertices belongs to $T$ conditioned on $A$ is
  \begin{align*}
    \Pr[Q_k \mid A]
    &\le \left(1-\frac{t}{2n}\right)^{n^z-s}
    \le \exp\left(-\frac{t}{2n}\left(n^z-s\right)\right).
  \end{align*}
  In total we get $\Pr[Q_k]\le \Pr[Q_k\mid A] + \Pr[\overline A]\le 2n^{-\log\log n} + e^{-\frac{t}{2n}(n^z-s)}$.
  
  It remains to account for our initial assumption that~$Y_{n-s-1}^S \leq \frac{\log n}{\sqrt{n}}$,
  which fails to hold with probability at most $n^{-\frac{\log{n}}{4}}$ by \cref{cor: two hop}.
  Thus, the overall probability that the foremost forest algorithm does not reach~$T$
  is at most $2n^{-\log\log n} + e^{-\frac{t}{2n}(n^z-s)} + n^{-\frac{\log{n}}{4}} \leq \frac{5}{2}n^{-\log\log n} + e^{-\frac{t}{2n}(n^z-s)}$.
\end{proof}

\section{Sharp Threshold for Giant Open Connected Component}
\label{sec:threshold}

In this section, we report on our first main result.

\begin{theorem}[Main Result for Open Components]\label{th:main}
	The function $\frac{\log n}{n}$ is a sharp threshold for Giant Open Connected Component. More specifically, there exists a function $\varepsilon(n) \in o\mathopen{}\left(\frac{\log n}{n}\right)$, such that the size of a largest open temporally connected component in $\G \in \F_{n,p}$ is
	\begin{enumerate}
			\item[(i)] $o(n)$ \aas, if $p<\frac{\log n}{n} - \varepsilon(n)$; and
			\item[(ii)] $n-o(n)$ \aas, if $p>\frac{\log n}{n} + \varepsilon(n)$.
	\end{enumerate}
\end{theorem}

We prove the lower bound on the threshold (i.e.~\cref{th:main} (i)) in \cref{sec:threshold-lower}.
The proof of this bound is a straightforward consequence of a result on foremost \emph{tree} growth in $\F_p(K_n)$ from \cite{CasteigtsRRZ21}.
The upper bound (i.e.~\cref{th:main} (ii)) on the threshold is proved in \cref{sec:threshold-upper}
and is significantly more involved.
In particular, it relies on~\cref{theorem:general target set} to measure foremost \emph{forest} growth in $\F_p(G)$, where $G$~is chosen to contain all edges that did \emph{not} occur within some particular time window.

\subsection{Lower Bound on the Threshold}
\label{sec:threshold-lower}
We state the lower bound in form of the following lemma which says that a.a.s.\ there is no linear size component before time~$\log n / n$.
This theorem can be derived rather easily from results in~\cite{CasteigtsRRZ21}.
\begin{lemma}[Lower Bound in~\cref{th:main}]\label{thm: lower bound}
    Let $\GGG \sim \FFF_{p}(K_n)$ with $p<\frac{\log n}{n} - \frac{3 (\log n)^{0.8}}{n}$. Then, for any constant $c\in (0,1)$, the graph $\GGG$ does not contain a temporally connected component of size at least $c\cdot n$ with probability at least $1-2n^{-\sqrt{\log n}}$.
\end{lemma}

We begin by reciting some relevant notation and results from~\cite{CasteigtsRRZ21}.
We observe that the foremost forest algorithm (Algorithm~\ref{alg: foremost forest}) applied to a singleton set $S= \{v\}$ produces a foremost tree rooted at $v$. In this case the algorithm coincides with the foremost tree algorithm from \cite{CasteigtsRRZ21}.
To state the relevant results about the growth of a foremost tree in $\F_p(K_n)$,
we first recapitulate the corresponding notation, which we will extend in \cref{sec: foremost forest analysis} to conduct the analysis of the foremost forest algorithm in the more general model of $\F_p(G)$.

Let $\GGG\sim\FFF_1(K_n)$ and let $v$ be a vertex in $\GGG$. The edges of the foremost
tree for $v$ computed by the algorithm are denoted as $e^v_{1},\ldots, e^v_{n-1}$ and
formally can be defined as follows. Let $\GGG_F^{0, v} :=(\{v\}, \emptyset, \emptyset)$. Then for every $k\in [n - 1]$
\begin{equation*}
	\begin{split}
		E^v_k 	&:= V(\GGG_F^{k - 1, v}) \times \big( V \setminus V(\GGG_F^{k - 1, v}) \big),\\
                e^v_{k}	&:= \arg\min \{ \lambda(e) ~|~ e \in E^v_k \text{ and } \GGG_F^{k - 1, v} \cup \{e \} \text{ is an increasing temporal forest} 
                \\ & \text{\hspace{10em} with $1$ component} \}, \\
		\GGG_F^{k, v} 	&:= \GGG_F^{k - 1, v} \cup \{ e^v_{k} \}.
	\end{split}
\end{equation*}

\noindent
Next, let
$Y^v_{0} := 0$ and $Y^v_k := \lambda(e^v_{k})$ for $k\in [n - 1]$.
By definition, for $k\in [n - 1]$, the random variable $Y^v_k$ is the earliest time when the foremost tree for $v$ contains exactly $k$ edges, or equivalently the earliest time by which $v$ can reach $k + 1$ vertices (itself included).
For $k\in [n - 1]$, let $X^v_k$ be a random variable equal to $Y^v_k - Y^v_{k-1}$,
i.e., to the \emph{waiting time} between the edges $e^v_{k-1}$ and $e^v_k$. Clearly, we have
$
Y^v_k = \sum_{i=1}^k X^v_i
$
for every $k \in [n - 1]$.
In addition, it is convenient to consider the \emph{truncated} versions of these variables.
For $k\in [n - 1]$, let
\[
\capped{X}^v_k := \min \{ X^v_k, c_k \} \,\text{ and }\,
\capped{Y}^v_k := \sum_{i=1}^{k} \capped{X}^v_i, \,\text{ where }\,
c_{k} := \frac{2 \log{(\min\{ k, n-k\})} + \log \log{n}}{k(n-k)}.
\]

With this notation at hand, we are now ready to state the technical results from \cite{CasteigtsRRZ21} that we are going to use later on.
\begin{theorem}[Theorem~4.8~\cite{CasteigtsRRZ21}]\label{thm: casteigts}
	With probability at least $1-2n^{-\sqrt{\log n}}$, for every vertex $v$ and $k\in[n-1]$, we have
	\[
	\left| \widehat{Y}_k^v - \sum_{i=1}^k \frac{1}{i(n-i)+1}\right|\le \frac{2 (\log n)^{0.8}}{n} .
	\]
\end{theorem}

\begin{lemma}[Lemma 4.12~\cite{CasteigtsRRZ21}]\label{lem: tree growth}
	For every function $z = z(n)$ with $0 \leq z(n) \leq 1$, and every $y > 0$ there is $n_0$ such that for all $n \geq n_0$
	a fixed vertex $v$ in $\G \sim \F_{p}(K_n)$
	can reach (resp.\ be reached by)
	\begin{enumerate}[(i)]
		\item \emph{at least} $\ceil*{\frac{n^z}{(\log n)^y}}$
		vertices with probability at least $1 - \frac{5}{\log{n}}$, if $p \geq z\frac{\log n}{n} + \frac{3(\log n)^{0.8}}{n}$;

		\item \emph{at most} $\ceil*{\frac{n^z}{(\log n)^y}}$
		vertices with probability at least $1 - \frac{5}{\log{n}}$, if $p \leq z\frac{\log n}{n} - \frac{3(\log n)^{0.8}}{n}$.
	\end{enumerate}
\end{lemma}

Now, the proof of~\cref{thm: lower bound} follows more or less immediately from the previously stated results from \cite{CasteigtsRRZ21}. We first note the following technicality which can be shown by an easy calculation similar to~\cref{cl: favsum} below, and which also follows from Lemma 4.7 in~\cite{CasteigtsRRZ21}: For every constant $c\in (0, 1)$,
\begin{align*}
    \Big|
        \sum_{i=1}^{\lceil c\cdot n \rceil} \frac{1}{i(n-i) + 1}
        - \frac{\log n}{n}
   \Big|
   \in \bigO\mathopen{}\left(\frac{1}{n}\right).
\end{align*}

This allows us to prove the following corollary.
\begin{corollary}\label{cor: cn}
  Let $c\in (0, 1)$ be any constant. With probability at least $1-2n^{-\sqrt{\log n}}$, for every vertex $v$, it holds that
  \[
    \left| \widehat{Y}_{\lceil c\cdot n \rceil}^v - \frac{\log n}{n}\right| \le \frac{3(\log n)^{0.8}}{n}.
  \]
\end{corollary}
\begin{proof}
    Using \cref{thm: casteigts} with $k=\lceil c\cdot n \rceil$ and the observation above,
    we conclude that, with probability at least $1-2n^{-\sqrt{\log n}}$, for every $v$ and for sufficiently large $n$, it holds that
    \begin{align*}
	    \left| \widehat{Y}_{\lceil c\cdot n \rceil}^v - \frac{\log n}{n}\right|
	    &=\left| \widehat{Y}_{\lceil c\cdot n \rceil}^v - \sum_{i=1}^{\lceil c\cdot n \rceil} \frac{1}{i(n-i)+1} + \sum_{i=1}^{\lceil c\cdot n \rceil} \frac{1}{i(n-i)+1}
	    - \frac{\log n}{n} \right|\\
	    &\le \left| \widehat{Y}_{\lceil c\cdot n \rceil}^v - \sum_{i=1}^{\lceil c\cdot n \rceil} \frac{1}{i(n-i)+1}\right|
	    + \left| \sum_{i=1}^{\lceil c\cdot n \rceil}\frac{1}{i(n-i) + 1} - \frac{\log n}{n} \right| \\
	    &\le\frac{2 (\log n)^{0.8}}{n}
	    + \bigO\mathopen{}\left(\frac{1}{n}\right)
	    \le \frac{3 (\log n)^{0.8}}{n}.\qedhere
    \end{align*}
\end{proof}
The corollary above together with the definition of the truncated variables is already enough to show that no node can reach a constant fraction of nodes before $\log n / n$. We are thus ready to prove~\cref{thm: lower bound}.
\begin{proof}[Proof of~\cref{thm: lower bound}]
    By definition of the truncated variables, as $\widehat{Y}_{k}^v\leq Y_{k}^v$ for every $k\in [n-1]$, it holds also that $\widehat{Y}_{c\cdot n}^v\leq Y_{\lceil c\cdot n \rceil}^v$.
    Using \cref{cor: cn}, with probability at least $1-2n^{-\sqrt{\log n}}$, for every vertex $v$, it holds that
    \[
        Y_{\lceil c\cdot n \rceil}^v
        \geq \widehat{Y}_{\lceil c\cdot n \rceil}^v
        \geq\frac{\log n}{n}-\frac{3(\log n)^{0.8}}{n}.
    \]
    Hence, with probability at least $1-2n^{-\sqrt{\log n}}$,
    no vertex can reach $\lceil c\cdot n \rceil$ nodes
    in $\GGG$ when $p<\frac{\log n}{n}-\frac{3(\log n)^{0.8}}{n}$ and, thus, cannot be in a temporally connected component of size at least $c\cdot n$. Hence, there is no temporally connected component of size at least $c\cdot n $ in $\GGG$.
\end{proof}

\subsection{Upper Bound on Threshold}
\label{sec:threshold-upper}
Next, we present the first, weaker version of our main result,
stating that an open temporally connected components containing almost all vertices appears already around time~$\log n / n$.

\begin{lemma}[Upper Bound in~\cref{th:main}]\label{thm: upper bound}
	Let $\GGG \sim \FFF_{p}(K_n)$ with $p\ge (1+\eps(n))\cdot \frac{\log n}{n}$. Then, the graph $\GGG$ contains a temporally connected component of size $n-o(n)$ a.a.s.
\end{lemma}

\begin{figure}[ht]
  \centering
  \resizebox{\textwidth}{!}{
\begin{tikzpicture}[x=1.8cm,y=2cm,scale=0.3, every node/.style={scale=0.3}]
\tikzset{
	mylabel/.style={scale=3},
	path/.style={decorate,decoration={snake,amplitude=1mm, segment length=15pt,pre length=2pt, post length=2pt},->,>=stealth'},
	mycloud/.style={thick,cloud, cloud puffs=25, cloud ignores aspect, minimum width=12cm, minimum height=16cm, draw},
	mybox/.style={minimum width=6cm,shape=rectangle,draw=black,line width=0.5mm},
}
\draw[line width=0.5mm] (-5,1)-- (25,1); 
\draw[line width=0.5mm] (-5, 0.8) -- (-5, 1.2) node[mylabel] (a) at (-5, 1.6) {$0$};
\draw[line width=0.5mm] (25, 0.8) -- (25, 1.2) node[mylabel] (a) at (24, 1.6) {$p=\frac{\log n}{n} + \eps$};
\draw[line width=0.5mm] (5, 0.8) -- (5, 1.2) node[mylabel] (a) at (5, 1.6) {$\frac{p}{3}$};
\draw[line width=0.5mm] (15, 0.8) -- (15, 1.2) node[mylabel] (a) at (15, 1.6) {$\frac{2p}{3}$};

\node[mylabel, color=red] (p1) at (-6.25, 0.6) {Phase 1};
\draw[line width=1mm, color=red] (-5, 0.6) -- (5, 0.6) node[mylabel, below, midway] {$I_1$};
\draw[line width=1mm, color=red] (15, 0.6) -- (25, 0.6) node[mylabel, below, midway] {$I_3$};
\node[mylabel, color=blue] (p1) at (-6.25, 0.1) {Phase 2};
\draw[line width=1mm, color=blue] (5, 0.1) -- (15, 0.1) node[mylabel, below, midway] {$I_2$};

\begin{scope}[yshift=1.5cm]
  \node[mycloud] (cloud) at (0, -6) {};
  \node[mybox, minimum height=6cm,fill=gray!20] (Zbox) at (0, -6.85) {};
  \draw[dashed] (-1.5, -7.2) -- (1.5, -7.2);
  \node[minimum width=4cm,shape=circle,draw=black,line width=0.5mm,fill=gray!20] (udisk) at (0,-3.5) {};
  \node[minimum size=0.1cm,fill=black, label={[mylabel]left:$x$}] (x) at (0,-6) {};
  \draw (x) -- (udisk.south west);
  \draw (x) -- (udisk.south east);
  \node[minimum size=0.1cm,fill=black, label={[mylabel]above:$u$}] (u) at (udisk) {};
  \draw[path] (x) -- (u);
  \node[mylabel] at (0,-6.6) {$Z=X\cap Y$};
  \node[mylabel] at (0,-7.75) {$X-Y$};
  \node[mylabel] at (0,-9.25) {$\G_{1}$};

  \draw[dashed] (5,-11) -- (5,0);

  \node[mycloud] (cloud) at (10, -6) {};
  \node[minimum width=4cm,shape=circle,draw=black,line width=0.5mm,fill=gray!20] (udisk) at (10,-3.5) {};
  \node[minimum size=0.1cm,fill=black, label={[mylabel]above:$u$}] (u) at (udisk) {};
  \node[mylabel] at (8,-6) {$\G_{2}$};
  \node[minimum width=4cm,shape=circle,draw=black,line width=0.5mm,fill=gray!20] (vdisk) at (10,-8.5) {};
  \node[minimum size=0.1cm,fill=black, label={[mylabel]below:$v$}] (v) at (vdisk) {};
  \draw[path] (u) -- (v);

  \draw[dashed] (15,-11) -- (15,0);

  \node[mycloud] (cloud) at (20, -6) {};
  \node[mylabel] at (20,-2.75) {$\G_{3}$};
  \node[mybox, minimum height=6cm,fill=gray!20] (Zbox) at (20, -5.5) {};
  \draw[dashed] (18.5, -5) -- (21.5, -5);
  \node[mylabel] at (20,-5.8) {$Z=X\cap Y$};
  \node[mylabel] at (20,-4.5) {$Y-X$};
  \node[minimum size=0.1cm,fill=black, label={[mylabel]right:$y$}] (y) at (20,-6.6) {};
  \node[minimum width=4cm,shape=circle,draw=black,line width=0.5mm,fill=gray!20] (vdisk) at (20,-8.5) {};
  \node[minimum size=0.1cm,fill=black, label={[mylabel]below:$v$}] (v) at (vdisk) {};
  \draw (y) -- (vdisk.north west);
  \draw (y) -- (vdisk.north east);
  \draw[path] (v) -- (y);
\end{scope}
\end{tikzpicture}
  }
  \caption{General strategy for upper bounding the value of $p$ in the case of open components.
    Here, $\G_{i}$ denotes the restriction of the temporal
    graph to subinterval $I_{i}$.
    Wavy lines denote temporal paths. We
    show that any node $x\in Z$ can reach any other node
    $y\in Z$ by reaching a node $u$ in $\G_{1}$, then a node $v$ in $\G_{2}$, and
    finally $y$ in $\G_{3}$. }
  \label{fig:proof_scheme}
\end{figure}
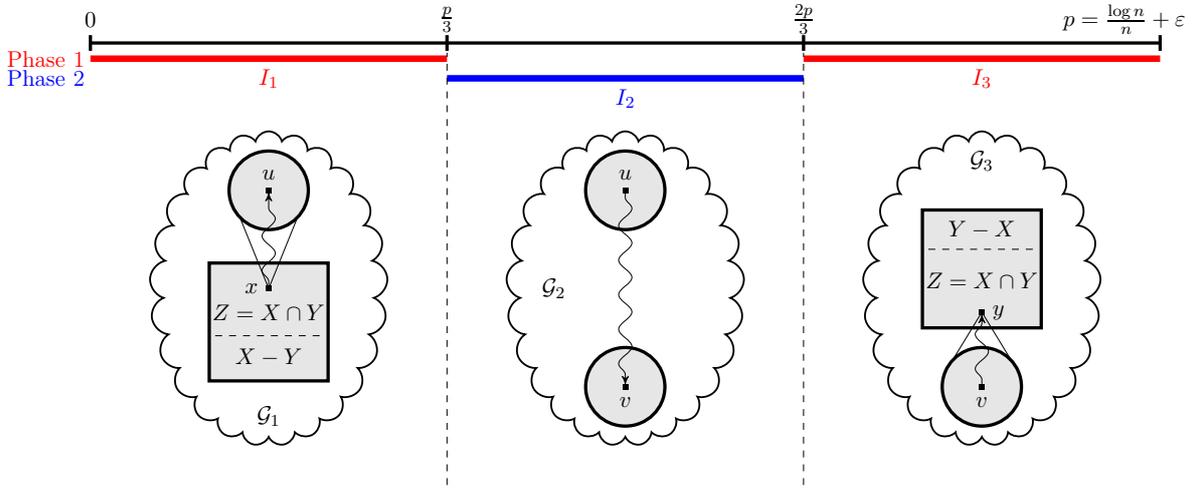

We begin by sketching the proof idea.
The strategy is as follows (see also \cref{fig:proof_scheme}). We split the time interval $[0, p]$
into three intervals $I_{1}$, $I_{2}$, and $I_{3}$ of equal duration $p / 3$, and
reveal the edges of the graph in two phases.

In Phase 1, we reveal the edges whose
time labels are in one of the intervals $I_1$ and $I_3$. Using a result from~\cite{CasteigtsRRZ21} 
(\cref{lem: tree growth}), we can conclude that there are
$n - o(n)$ nodes (call them $X$), each of which \aas reaches at
least $\sqrt[3]{n}\log n$ vertices during $I_1$, and there are at least
$n - o(n)$ nodes (call them $Y$) that \aas is reached by at least $\sqrt[3]{n}\log n$
vertices during $I_3$.

In Phase 2, we reveal the edges appearing during the middle interval $I_2$.
We show that for every ordered pair of nodes $x,y$ in the set $Z := X \cap Y$ 
(which is our intended connected component), the set of vertices that $x$ can reach during $I_1$,
can reach during $I_2$ at least one vertex in the set of vertices that reach $y$ during $I_3$; thus
implying that $x$ can reach $y$ during $[0,p]$. For this purpose we can employ Theorem~\ref{theorem:general target set} with $S$ being the set that $x$ can reach during $I_1$ and $T$ being the set of vertices that can reach $y$ during $I_3$. Note that the analysis of this phase is what requires us to develop the generalization $\F_p(G)$ of the model $\F_{n,p}$. In fact, the static base graph $G$ used in the application of Theorem~\ref{theorem:general target set} is the graph obtained from $K_n$ by removing the edges that appeared during either $I_1$ or $I_3$. Finally, a union bound over all pairs of nodes $x$ and $y$ yields the result.

The remainder of this section is dedicated to proving~\cref{thm: upper bound} formally.
Throughout, we denote $\eps(n):=\frac{1}{\log\log n}$.
Let $p=(1+\eps(n))\cdot \frac{\log n}{n}$ and $\GGG \sim \FFF_{p}(K_n)$.
We will prove~\cref{thm: upper bound} only for this value of $p$ as
it will then clearly follow for any larger value.
Our strategy is to split the interval $[0, p]$ into three sub-intervals $[p_0,p_1]$, $[p_1,p_2]$, $[p_2,p_3]$, where $p_i:=\frac{i}{3}(1+\eps(n))\frac{\log n}{n}$ for $i\in[0, 3]$.
We first deduce the following corollary about the connectivity of the subgraphs $\GGG_{[p_i, p_{i+1}]}$ for $i\in [0, 2]$ of $\GGG$ from~\cref{lem: tree growth}.

\begin{corollary}\label{cor: reach thirds}
  For $i\in [0, 2]$, the number of vertices \emph{reached by} (resp. \emph{reaching}) a fixed vertex in
  $\GGG_{[p_i,p_{i+1}]}$ lies within $[n^{1/3}\log n, n^{1/3 + \eps(n)}]$ with probability at least $1 - \frac{10}{\log n}$.
\end{corollary}
\begin{proof}
		We show the statement for $\GGG_{[0, p_1]}$, the other statements follow analogously via shifting the edge labels.
        We start with the lower bound. We apply~\cref{lem: tree growth}~(i) with $z(n)=\frac{1}{3} + \frac{2}{\eps(n)\log n}$ and $y=1$. We get that for large enough $n$ a fixed vertex in $\GGG\sim\FFF_{p}(K_n)$ can reach (resp. be reached by) at least
        \[
            \ceil*{\frac{n^z}{(\log n)^y}}
						\ge \frac{n^{\frac{1}{3} + \frac{2\log \log n}{\log n}}}{\log n}
            = n^{\frac{1}{3}} \log n
        \]
        nodes with probability at least $1-\frac{5}{\log n}$, if $p\ge p'(n)$, where $p'(n):=z \frac{\log n}{n} + 3 \frac{(\log n)^{0.8}}{n}$. Observe that
        \[
            p_1 \cdot n
            = \frac{1}{3}\cdot \log n + \frac{\eps(n)}{3} \cdot \log n
            \ge \frac{1}{3}\cdot \log n + \frac{2}{\eps(n)} + 3(\log n)^{0.8}
						= p'(n) \cdot n
        \]
				for sufficiently large $n$.

				In order to prove the upper bound, we apply~\cref{lem: tree growth}~(ii) with $z(n)=\frac{1}{3} + \eps(n)$ and $y=1$. We get that for large enough $n$ a fixed vertex in $\GGG\sim\FFF_{p}(K_n)$ can reach (resp. be reached by) at most
        \[
            \ceil*{\frac{n^z}{(\log n)^y}}
						\le \frac{n^{\frac{1}{3} + \eps(n)}}{\log n} + 1
						\le n^{\frac{1}{3} + \eps(n)}
        \]
        nodes with probability at least $1-\frac{5}{\log n}$, if $p\le p'(n)$, where $p'(n):=z \frac{\log n}{n} - 3 \frac{(\log n)^{0.8}}{n}$. Observe that
        \[
            p_1 \cdot n
            = \Big(\frac{1}{3} + \eps(n)\Big) \cdot \log n - \frac{2\eps(n)}{3} \cdot\log n
            \le \Big(\frac{1}{3} + \eps(n)\Big)\cdot  \log n - 3(\log n)^{0.8}
						= p'(n) \cdot n
        \]
				for sufficiently large $n$.

				We conclude by union bound that the cardinality of the set of vertices that a fixed vertex reaches (resp. is reached by) in $\GGG_{[0, p_1]}$ is in the wanted interval with probability at least $1-\frac{10}{\log n}$.
\end{proof}

Using Markov's inequality we can obtain that, a.a.s., almost all nodes can reach (resp. be reached by) the above number of nodes.
\begin{lemma}
\label{lemma: almost everyone reaches many}
				Let $i\in \{0, 1, 2\}$. The number of vertices that can reach (resp. be reached by) at least $n^{1/3}\log n$ and at most $n^{1/3 + \eps(n)}$ vertices in $\GGG_{[p_i, p_{i+1}]}$ is at least $n - \frac{n}{\log \log n}$ with probability at least $1 - \frac{10 \log \log n}{\log n}$.
\end{lemma}
\begin{proof}
				Let $\bar X$ denote the number of nodes in $\GGG_{[p_i, p_{i+1}]}$ that can reach (resp. be reached by) less than $n^{1/3}\log n$ or more than $n^{1/3 + \eps(n)}$ vertices in $\GGG_{[p_i, p_{i+1}]}$. Then $\E[\bar X] \le 10n /\log n$ by~\cref{cor: reach thirds}. Using Markov's inequality
        $
            \Pr\Big[\bar X \geq \frac{n}{\log\log n}\Big]
            \le \frac{10 \log \log n}{\log n}.
        $
\end{proof}
We now denote by $X$ the set of nodes that can reach at least $n^{1/3}\log n$ and at most $n^{1/3 + \eps(n)}$ vertices in $\GGG_{[0, p_{1}]}$ and by $Y$ the set of nodes that are reached by at least $n^{1/3}\log n$ and at most $n^{1/3 + \eps(n)}$ vertices in $\GGG_{[p_2, p_{3}]}$. Furthermore, we denote by $Z=X\cap Y$ their intersection.
According to~\cref{lemma: almost everyone reaches many}, it holds that $|Z|\ge n - \frac{2n}{\log \log n}$ with probability at least $1 - \frac{20 \log \log n}{\log n}$.
The hardest part of our proof is to now show that, for a fixed ordered pair $x,y\in Z$, the probability that there is a temporal path from $x$ to $y$ is so large that we can take a union bound over all ordered pairs. To this end, let $A(x)$ be the set of nodes that $x$ can reach in $\GGG_{[0,p_1]}$ and let $B(y)$ be the set of nodes that can reach $y$ in $\GGG_{[p_2, p_3]}$. Furthermore, for $x\in X$, let
\[
    A'(x):=\{v\in V: \exists a\in A(x) \text{ s.t.\ }a \text{ reaches } v \text{ in }\GGG_{[p_1, p_2]}\}
\]
be the set of nodes that $x$ can reach in $\GGG_{[0, p_2]}$.
Notice that $x$ reaches $y$ if and only if $A'(x)$ intersects $B(y)$.

Let $G'' = (V, E'')$ with
$E'' = \{e \in \binom{V}{2} \mid \lambda(e) \in [0, p_1] \cup [p_2, p_3]\}$
be the graph containing all edges appearing in $\GGG_{[0, p_1]}$ or $\GGG_{[p_2, p_3]}$,
and let $G' = (V, E')$ with $E' = \binom{V}{2} \setminus E''$ contain all other edges.
Then we observe that the distribution of the set $A'(x)$ conditioned on the information about the edges appearing in $\GGG_{[0,p_1]}$ and $\GGG_{[p_2, p_3]}$ is identical to the node set of a foremost forest grown from $S:=A(x)$ in $\HHH\sim \FFF_{p'}(G')$,
where $p'=\frac{1}{3}(1+\eps(n))\frac{\log n}{n}$.
Furthermore, $G''$ is distributed as an \ER{} graph $G'' \sim G_{n, p}$ with $p:=\frac{2}{3}(1+\eps(n))\frac{\log n}{n}$.
From a standard result regarding the maximum degree in $G_{n,p}$ we can thus conclude the following fact.
\begin{observation}\label{obs: min degree}
	It holds that $\Delta(G'')\le 4\log n$ a.a.s.\ and, thus, $\delta(G')\ge n - (\log n)^2$ a.a.s.
\end{observation}
\begin{proof}
	Recall that $G''$ is distributed according to $G_{n, p}$ with $p:=\frac{2}{3}(1+\eps(n))\frac{\log n}{n}$. Following~\cite[Corollary 3.13]{Bollobas2001}, with $m=1$ and $\omega(n)=\log n$, we have that a.a.s.\
	\[
			\Delta(G'') \leq pn + \sqrt{2pn\log n} + \log n \sqrt{\frac{pn}{\log n}}
			\le \log n + \sqrt{2 (\log n)^2} + \log n
			\le 4 \log n.
	\]
	The observation about the minimum degree now follows immediately for sufficiently large $n$.
\end{proof}
Thus, in order to lower bound the probability that $A'(x)$ intersects $B(y)$, we can use the following corollary of~\cref{theorem:general target set}.
\begin{corollary}\label{corollary: target set}
  Let
  \begin{itemize}
    \item $G$ be a graph of minimum degree $\delta(G)\ge n - (\log n)^a$ for some $a\in\NN$,
    \item let $S$ and $T$ be two sets of nodes in $G$, each of cardinality at least $n^{1/3}\log n$, and
    \item let $\GGG\sim \FFF_{p}(G)$ with $p\ge \frac{1}{3}(1+\eps(n))\frac{\log n}{n}$.
  \end{itemize}
  Then, the foremost forest algorithm from $S$ on $\GGG$ reaches $T$ with probability at least $1-3n^{-\log\log n}$.
\end{corollary}
\begin{proof}
	Set $s := \abs{S}$, $t := \abs{T}$.
	Without loss of generality, we may assume $s \leq n^{1/3 + \eps(n)}$.
	Note that for large enough $n$ it holds that
	\begin{align*}
		p
    &\ge \frac{1}{3}\Big(1+\frac{1}{\log \log n}\Big)\frac{\log n}{n}\\
    &\ge \frac{\frac{1}{3}\log n + 4 \log \log n}{n}\\
    &= \frac{\frac{2}{3}\log n + 2 \log \log n - \frac{1}{3}\log n - \log \log n}{n} + \frac{3\log \log n}{n}\\
    &\ge \frac{z\log n - \log s}{n} + \frac{3\log \log n}{n},
	\end{align*}
  for $z = \frac{2}{3} + \frac{2\log\log n}{\log n}$.
  From \cref{theorem:general target set} it then follows that the foremost forest algorithm from $S$ reaches $T$ with probability at least
  \begin{align*}
    1 - \frac{5}{2} n^{-\log \log n} - e^{-\frac{t}{2n}(n^z - s)}
    &\ge 1 -  \frac{5}{2} n^{-\log \log n} - e^{-\frac{n^{1/3}\log n}{2n}(n^{2/3}(\log n)^2 - n^{1/3 + \eps(n))})}\\
    &\ge 1 -  \frac{5}{2} n^{-\log \log n} - e^{-\frac{(\log n)^3}{4}}
    \ge 1 -3 n^{-\log \log n},
  \end{align*}
  completing the proof.
\end{proof}

Using the above stated corollary, we can finally prove our first main result.
\begin{proof}[Proof of~\cref{thm: upper bound}]
	Let $p= (1+\eps(n))\cdot \frac{\log n}{n}$ and $\GGG \sim \FFF_{p}(K_n)$. As above, let $X$ be the nodes that can reach between $n^{1/3}\log n$ and $n^{1/3 + \eps(n)}$ vertices in $\GGG_{[0, p_{1}]}$ and let $Y$ be the nodes that are reached by between $n^{1/3}\log n$ and $n^{1/3 + \eps(n)}$ vertices in $\GGG_{[p_2, p_{3}]}$.
	Furthermore, let $Z=X\cap Y$ be their intersection and recall that $|Z|\ge n - \frac{2n}{\log \log n}$ with probability at least $1 - \frac{20 \log \log n}{\log n}$ according to~\cref{lemma: almost everyone reaches many}.
	Now, conditioned on the information about the edges appearing in $\GGG_{[0,p_1]}$ and $\GGG_{[p_2, p_3]}$, let $G'=(V, E')$ be the static graph with the same node set as $\GGG$ and the edge set $E'=\{e\in \binom{V}{2}:\lambda(e)\notin [p_0, p_1]\cup [p_2, p_3]\}$, where $\lambda$ is the time label function of $\GGG$.
	Note that according to~\cref{obs: min degree} the minimum degree in $G'$ a.a.s.\ is at least $n - (\log n)^2$.
	Now, let $x,y\in Z$ be a fixed ordered pair of vertices.
	Applying~\cref{corollary: target set} to $\HHH\sim\FFF_{\frac{p}{3}}(G')$ with $S=A(x)$, $a=2$, and $T=B(y)$, we can conclude that $A'(x)\cap B(y)\neq \emptyset$ with probability at least $1 - 3n^{-\log\log n}$, and, thus, $x$ reaches $y$ with at least that probability.
	Hence, after a union bound over all ordered pairs, we get that all nodes in $Z$ reach each other with probability at least $1 - 3n^{-\log\log n + 2}$. Therefore, $\GGG$ has a temporally connected component of size at least $n - \frac{2n}{\log\log n} = n - o(n)$ a.a.s.
	\end{proof}

Thus both parts \cref{th:main} are now proven.

\section{Sharp Threshold for Giant Closed Connected Component}
\label{sec:threshold-closed}

In this section we prove that $\frac{\log n}{n}$ is also a sharp threshold for the existence of a giant \emph{closed} connected component.

\begin{theorem}[Main Result for Closed Components]
	\label{thm:closed-component}
	The function $\frac{\log n}{n}$ is a sharp threshold for Giant Closed Connected Component.
	More precisely, there exists a function $\eps(n) \in o\pfrac{\log n}{n}$,
	such that the size of a largest closed temporally connected component in~$\G \sim \F_{n,p}$ is
	\begin{enumerate}[(i)]
		\item $o(n)$ a.a.s., if $p < \frac{\log n}{n} - \eps(n)$; and
		\item $n-o(n)$ a.a.s., if $p > \frac{\log n}{n} + \eps(n)$.
	\end{enumerate}
\end{theorem}

We start by sketching the general proof idea.
  The lower bound of \cref{thm:closed-component} is obviously a trivial consequence of the lower bound in~\cref{th:main}.
  Thus, it remains to prove the upper bound.
  We employ our strategy of splitting the time into three intervals. 
  We do not need to make any changes to our approach in the middle one (Phase 2), which previously required the most effort.
  However, we now need to do additional work in the first and last interval (Phase 1), which is the main technical contribution of this part.
  Recall that in the proof of \cref{thm: upper bound}, we only required that $n-o(n)$~vertices can all reach (resp.\ be reached by) at least $n^{1/3} \log n$~vertices within each of the three intervals from Figure~\ref{fig:proof_scheme}.
  Now, we will need to prove that there exists a set~$V'$ of $n-o(n)$~vertices,
  such that every vertex in this set can reach (resp.\ be reached by) at least $n^{1/3}\log n$~vertices \emph{via temporal paths that use only vertices in~$V'$}. Once this is achieved, we can use the same approach as in the case of open components for Phase 2.

\begin{figure}[bt]
  \begin{center}
  \resizebox{\textwidth}{!}{
\begin{tikzpicture}[x=1.8cm,y=2cm,scale=0.3, every node/.style={scale=0.3}]
  \tikzset{mylabel/.style={scale=1.3}}
  
  \draw[line width=0.2mm] (0,0)-- (11,0); 
  \draw[line width=0.2mm] (0,0.1) -- (0,-0.1) node[mylabel] (a) at (0,0.3) {$0$};
  \draw[line width=0.2mm] (0.5,0.1) -- (0.5,-0.1) node[mylabel] (a) at (1.0,0.3) {$\eps_1= o\big(\frac{\log n}{n}\big)$};
  \draw[line width=0.2mm] (3.75,0.1) -- (3.75,-0.1) node[mylabel] (a) at (3.75,0.3) {$p_1 + \eps_1$};
  \draw[line width=0.2mm] (7.25,0.1) -- (7.25,-0.1) node[mylabel] (a) at (7.25,0.3) {$p - (p_1 + \eps_1)$};
  \draw[line width=0.2mm] (10.5,0.1) -- (10.5,-0.1) node[mylabel] (a) at (10.3, 0.3) {$p - \eps_1$};
  \draw[line width=0.2mm] (11,0.1) -- (11,-0.1) node[mylabel] (a) at (11.5, 0.3) {$p=\frac{\log n}{n} + \eps 
  $};
  
  \draw[line width=0.5mm, color=pink] (0,-0.21) -- (0.5,-0.21) node[mylabel, below, midway] {$I_1$};
  \draw[line width=0.5mm, color=pink] (10.5,-0.21) -- (11,-0.21) node[mylabel, below, midway] {$I_5$};
  \node[mylabel, color=pink] (p1) at (-0.7,-0.21) {Phase 1.1};

  \draw[line width=0.5mm, color=red] (0.5,-0.42) -- (3.75,-0.42) node[mylabel, below, midway] {$I_2$};
  \draw[line width=0.5mm, color=red] (7.25,-0.42) -- (10.5,-0.42) node[mylabel, below, midway] {$I_4$};
  \node[mylabel, color=red] (p1) at (-0.7,-0.42) {Phase 1.2};
  
  \draw[line width=0.5mm, color=blue] (3.75,-0.63) -- (7.25,-0.63) node[mylabel, below, midway] {$I_3$};
  \node[mylabel, color=blue] (p1) at (-0.8,-0.63) {Phase 2};
  
  \draw[dotted, line width=0.15mm] (0,0) -- (0, -0.21);
  \draw[dotted, line width=0.15mm] (11,0) -- (11, -0.21);
  \draw[dotted, line width=0.15mm] (0.5,0) -- (0.5, -0.42);
  \draw[dotted, line width=0.15mm] (10.5,0) -- (10.5, -0.42);
  \draw[dotted, line width=0.15mm] (3.75,0) -- (3.75, -0.63);
  \draw[dotted, line width=0.15mm] (7.25,0) -- (7.25, -0.63);
\end{tikzpicture}
  }
  \end{center}
  \caption{Illustration of the three different phases in our proof for the case of closed components. Here, the length $p_1$ of $I_2$ and $I_4$ and the length $p_2$ of $I_3$ are each roughly $\frac{1}{3}\frac{\log n}{n}$.
  In Phase 1.1, we reveal edges in $I_1$ and $I_5$ and identify our target closed connected component, a set of $n'$ nodes $V'$ each of which can reach (be reached by) poly-logarithmically many vertices within $V'$ during $I_1$ ($I_5$) via temporal paths in $V'$. 
  For Phase 1.2 (consisting of intervals $I_2$ and $I_4$) we show that every vertex in $v \in V'$ reaches (is reached by) polynomially many vertices in $V'$ during $I_1 \cup I_2$ ($I_4 \cup I_5$). 
  We then show that during Phase 2 (consisting of $I_3$), for each ordered pair of vertices $u,w \in V'$, the set of vertices reached by $u$ during $I_1 \cup I_2$ can reach the set of vertices that reach $w$ during $I_4 \cup I_5$, implying that $u$ can reach $w$ during $[0,p]$.}
  \label{fig:closed_component_three_phases}
\end{figure}
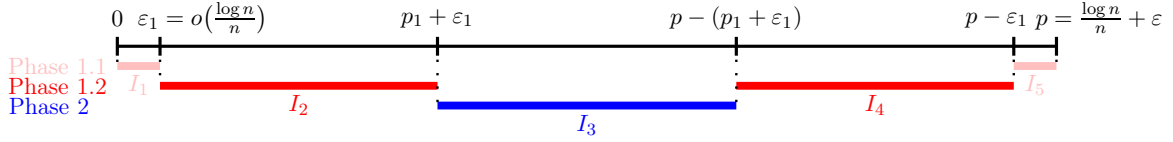

  In order to obtain the set of vertices $V'$ mentioned above, we have to insert an additional Phase 1.1,
  which looks only at a short time interval~$I_1$ at the very beginning (and symmetrically $I_5$ at the very end).
  The purpose of this new phase is to ``bootstrap'' the closed component by identifying a set~$V'$ of $n' = n - o(n)$~vertices,
  which each reach at least \emph{poly-logarithmically} many vertices by paths that are contained in~$V'$.
  \Cref{lm:bootstrap-reachability} formalizes this result.
  
\begin{figure}[hbt]
  \centering
\begin{tikzpicture}[scale=2]
	\draw[border]
		(190:1) arc[radius=1cm, start angle=190, end angle=-70] (-70:1)
		(0,0) -- (180:1)
		(0,0) -- (120:1)
		(0,0) -- (60:1)
		(0,0) -- (0:1)
		(0,0) -- (-60:1)
		;
	
	\path
		\foreach \x in {-70,-80,-90} {
			(\x:0.7) node[circle,fill=black,minimum size=0.8mm,inner sep=0] {}
		};
	
	\path
		(180-30:1.2) node {$V_i$}
		\foreach \i in {1,2,3} {
			(180-30-\i*60:1.2) node {$V_{i + \i}$}
		};
	
	\draw[edge]
		(140:0.6) node[vertex,label=left:$v$] (v) {}
		(v) -- (80:0.9) node[vertex] (v1) {}
		(v) -- (100:0.5) node[vertex] (v2) {}
		(v1) -- (20:0.9) node[vertex] (v11) {}
		(v1) -- (40:0.7) node[vertex] (v12) {}
		(v2) -- (30:0.5) node[vertex] (v21) {}
		(v2) -- (40:0.3) node[vertex] (v22) {}
		(v11) -- (-10:0.9) node[vertex] {}
		(v11) -- (-30:0.9) node[vertex] {}
		(v12) -- (-20:0.75) node[vertex] {}
		(v12) -- (-50:0.8) node[vertex] {}
		(v21) -- (-50:0.6) node[vertex] {}
		(v21) -- (-30:0.7) node[vertex] {}
		(v22) -- (-30:0.4) node[vertex] {}
		(v22) -- (-20:0.2) node[vertex] {}
		;
\end{tikzpicture}
  \caption{
    Example of a temporal tree formed by clocked paths starting at a vertex~$v \in V_i$.
    By restricting the edges used between sectors~$V_{i+j}$ and~$V_{i+j+1}$ to an appropriate time interval $I_{j-i}$,
    we ensure that the time labels of all these paths are monotonically clockwise increasing.
  }
  \label{fig:sectors}
\end{figure}
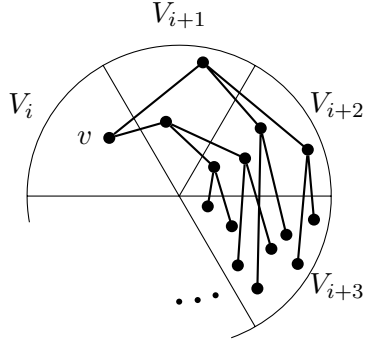

  A technical difficulty in Phase 1.1 is the need to control possible cascading effects,
  where removing low-degree vertices from the graph can cause further vertices to become low-degree vertices, etc.
  We overcome this difficulty by partitioning the vertices into sets $V_1, \dots, V_C$, which we refer to as \emph{sectors}, and removing vertices from each sector~$V_i$ solely on the base of whether they have too few neighbors in the \emph{next sector}~$V_{i+1}$.
  This ensures that the sets of vertices removed from each sector are determined independently of each other.
  On this base, we are then able to prove that no cascading effects occur a.a.s.
  Subsequently, we show that after these removals, every remaining vertex can reach a poly-logarithmic number of others by considering \emph{clocked} paths, which essentially march in lockstep, traversing the sectors in circular order (see \cref{fig:sectors}).
  
  Subsequently, in Phase 1.2, we reveal edges that appear during $I_2$ or $I_4$. 
  We use the foremost forest technique developed earlier to show that,
  conditioned on the edges revealed in Phase 1.1, for every vertex $v$ in $V'$ the poly-logarithmic set of
  vertices reached by~$v$ during~$I_1$ reaches \emph{polynomially many} (by which we mean $n^p$ for some fixed $p<1$) vertices during~$I_2$. (Similarly,
  the set of vertices that reach~$v$ during~$I_5$ is reached by polynomially many 
  vertices during $I_4$.)

We now proceed with the formal proof of \cref{thm:closed-component}, which is based on the following lemma, the proof of which is deferred until \cref{sec:ProofBootstrap}.

\begin{restatable}{lemma}{bootstrapReachability}
	\label{lm:bootstrap-reachability}
	Let $C \geq 3$, $\frac{1}{2}<\gamma<\alpha<1$, and let
	$\G \sim \F_{n,p}$, where~$p = 2C^2 \frac{(\log n)^\alpha}{n}$.
	Then \aas $\G$ contains a set~$V'$ of $n-o(n)$~vertices,
	such that, denoting $\G' := \G[V']$, every vertex in $V'$ 
	reaches at least $(\log n)^{(C-3)\gamma}$~vertices in $\G'_{[0,p/2]}$ and
	is reached by at least the same number of vertices in $\G'_{[p/2,p]}$.
\end{restatable}

\begin{proof}[Proof of \cref{thm:closed-component}]
	Let $\gamma=0.7$, $\alpha=0.9$, $C = 30$,
	and let $n'(n) \in n-o(n)$ be the size of the vertex set guaranteed by \cref{lm:bootstrap-reachability}.
	Set $\eps_1 := C^2 \frac{(\log n)^\alpha}{n} \in o\pfrac{\log n}{n}$,
	$\eps_2 := 4\frac{\log \log n'}{\log n'} \in o(1)$, and
	$\eps_3 := \frac{1}{3\log \log n} \in o(1)$.
	Set also $p_1 := \left(\frac{1}{3} + \eps_2\right) \frac{\log n'}{n'}$ and 
	$p_2 := \left(\frac{1}{3} + \eps_3\right) \frac{\log n'}{n'}$.
	Finally, define 
	$
	p := 2\eps_1 + 2p_1 + p_2,
	$
	which is equal to $\frac{\log n}{n} + \eps$ for some $\eps \in o\pfrac{\log n}{n}$.

	Let $\G \sim \F_{n,p}$. We split $[0,p]$
	into a total of five intervals $I_i$, $i \in [5]$. The first and the last interval are short and each has length $\eps_1$, i.e., $I_1 = [0, \eps_1]$ and $I_5 = [p-\eps_1, p]$. The three middle intervals are long
	and have lengths $p_1, p_2$, and $p_1$, respectively, i.e., $I_2=[\eps_1, \eps_1 + p_1]$,
	$I_3 = [\eps_1 + p_1, \eps_1 + p_1 + p_2]$, $I_4 = [p - (\eps_1 + p_1), p-\eps_1]$.
	We will reveal the edges of the graph in three phases (Phase 1.1, Phase, 1.2, and Phase 2), as was graphically summarized in \cref{fig:closed_component_three_phases} in the introduction, and in each phase we condition on the edges revealed in the previous phases. In Phase 1.1 we reveal edges in the intervals $I_1$ and
	$I_5$ and apply \cref{lm:bootstrap-reachability} to identify a large set of nodes $V'$, each of which can reach poly-logarithmically many vertices in $V'$ during the first interval and can be reached by poly-logarithmically many vertices in $V'$ during the last interval via temporal paths that use only nodes from $V'$. In the subsequent phases we restrict our attention to the subgraph induced by $V'$, which is the target giant closed connected component.
	In Phase 1.2, we reveal edges appearing in the intervals $I_2$ and $I_4$. Because in 
	Phase 1.1 \aas we revealed poly-logarithmic number of edges for every vertex, we can use \cref{lem: reachability} to argue that for every vertex $v \in V'$ the poly-logarithmic set of vertices reached by $v$ during $I_1$ can reach polynomially many vertices during $I_2$. Similarly, during $I_4$ polynomially many
	vertices can reach the poly-logarithmic set of vertices that reach $v$ during $I_5$. The main outcome of this phase is that every vertex in $v \in V'$ reaches polynomially many vertices in $V'$ during $I_1 \cup I_2$ and is reached by at least as many vertices in $I_4 \cup I_5$. 
	Finally, in Phase 2, because in the previous phases \aas at most poly-logarithmically many
	edges were revealed for every vertex, we can apply \cref{corollary: target set} to prove that for each ordered pair of vertices $u,w \in V'$ the set of vertices reached by $u$ during $I_1 \cup I_2$ can reach during $I_3$ the set of vertices that reach $w$ during $I_4 \cup I_5$, implying that $u$ can reach $w$ during $[0,p]$. We now proceed with the formal details.

	\paragraph*{Phase 1.1.}
	Let $\G_1$ be the temporal subgraph of $\G$ formed by the edges with time labels in the intervals $I_1 \cup I_5$. Note that, up to shifting the time labels in the interval $I_5$ by
	$p - 2\eps_1$, $\G_1$ is distributed according to $\F_{n, 2\eps_1}$.
	Thus, by \cref{lm:bootstrap-reachability}, \aas there is a set $V' \subseteq V(\G)$ containing
	$n'$ vertices such that, denoting $\G' := \G[V']$, every vertex $v \in V'$ 
	reaches a set $R_1(v)$ of at least $(\log n)^{(C-3)\gamma}$~vertices in $\G'_{I_1}$ and
	is reached by a set $R_1'(v)$ of at least $(\log n)^{(C-3)\gamma}$~vertices in $\G'_{I_5}$.
	
	\paragraph*{Phase 1.2.}
	Let $G_1$ be the underlying graph of $\G_1$. 
	Since $G_1$ is distributed as an \ER{} graph $G_{n,\,2\eps_1}$,
	similarly to \cref{obs: min degree}, we have that $\Delta(G_1) < 4 \log n$ a.a.s.
	Hence, in the graph $G_2' = \left(V', {V' \choose 2} \setminus E(G_1) \right)$ the minimum degree is at least $n' - 4 \log n \geq n' - (\log n')^2$.
	Observe that, up to shifting time labels, $\G'_{I_2} \sim \F_{p_1}(G_2')$ when conditioning on the knowledge about all edges seen in $I_1 \cup I_5$.
	Therefore, since $\abs{R_1(v)} \geq (\log n')^{13}$ for every vertex $v \in V'$, by applying
	\cref{lem: reachability} to $\G'_{I_2}$ and $R_1(v)$ 
	(with parameter $z = 1/3 + \frac{\log \log n'}{\log n'}$), we conclude that
	the vertices in $R_1(v)$ reach in $\G'_{I_2}$ at least $r := {(n')}^{\frac{1}{3} + \frac{\log\log n'}{\log n'}} = (n')^{1/3} \log n'$ vertices with probability at least $1-2(n')^{-\log\log n'}$.
	By the union bound, we have that with probability at least $1-2(n')^{1-\log\log n'} \in 1-o(1)$, every vertex $v \in V'$ can reach in $\G'_{I_1 \cup I_2}$ a set $R_2(v)$ of at least $r$ vertices.
	Symmetrically, with probability at least $1-2(n')^{1-\log\log n'} \in 1-o(1)$, every vertex $v \in V'$ is reached in $\G'_{I_4 \cup I_5}$ by a set~$R_2'(v)$ of at least $r$ vertices.
	
	\paragraph*{Phase 2.}
	Let $G_3'$ be the static graph defined by the vertex set $V'$ and all edges appearing in $\G'_{I_1 \cup I_2}$ and $\G'_{I_4 \cup I_5}$. As in Phase 1.2, we can argue that the maximum degree of $G_3'$ is at most $4 \log n'$ \aas, and therefore the minimum degree of the graph $G_4' = \left(V', {V' \choose 2} \setminus E(G_3') \right)$ 
	is at least $n' - (\log n')^2$. Hence, up to shifting time labels, $\G'_{I_3}$ is distributed according to $\F_{p_2}(G_4')$ when conditioned on the knowledge of all edges revealed in $I_1 \cup I_2 \cup I_4 \cup I_5$.
	Thus, by \cref{corollary: target set}, a given set of at least $(n')^{1/3} \log n'$ vertices in $\G'_{I_3}$ can reach another given set of at least as many vertices with probability at least $1-3(n')^{-\log\log n'}$. Applying this to all ordered pairs of sets $(R_2(v), R'_2(w))$, $v, w \in V'$ and
	using the union bound, we conclude that
	the probability that all these pairs of sets reach each other in $\G'_{I_3}$ is at least $1-3(n')^{2-\log\log n'} \in 1-o(1)$.
	
	Putting all together, we conclude that \aas in $\G' = \G[V']$ any vertex reaches every other vertex.
	Thus, $V'$ is, as desired, a giant closed connected component.
\end{proof}

\subsection{Proof of \cref{lm:bootstrap-reachability}}
\label{sec:ProofBootstrap}
In what follows we will make repeated use of Chernoff bounds. For better readability, we summarize the different versions that we use in the following lemma (see, e.g., Theorems~4.4 and 4.5 in~\cite{MitzenmacherU17}).
\begin{lemma}[Chernoff bounds]\label{lem: chernoff}
  Let $X_1,\ldots, X_T$ be independent random variables taking values in $\{0,1\}$, let $X$ denote their sum, and let $\mu=\E[X]$. Then, the following hold:
  \begin{enumerate}[(i)]
    \item\label{chernoff 1} 
    For any $\delta>0$,
    \[
      \Pr[X\ge (1+\delta) \mu] 
      \le \Big( \frac{e^\delta}{(1+\delta)^{1+\delta}} \Big)^\mu
      \le \exp\Big(-\frac{\mu \delta^2}{2+\delta}\Big).
    \]
    \item\label{chernoff 2}
    For any $R\ge 6\mu$,
    \[
      \Pr[X\ge R] \le 2^{-R}.
    \]
    \item\label{chernoff 3}
    For any $0<\delta<1$,
    \[
      \Pr[X\le (1-\delta) \mu] \le \exp\Big(-\frac{\mu \delta^2}{2}\Big).
    \]
  \end{enumerate} 
\end{lemma}

We proceed with an informal description of the proof strategy.
Recall, that we would like to show that for any fixed integer $C \geq 3$ and constants $\frac{1}{2}<\gamma<\alpha<1$, a random temporal graph
$\G \sim \F_{n,p}$ with $p = 2C^2 \frac{(\log n)^\alpha}{n}$
\aas contains a set $V'$ of $n-o(n)$ vertices with the property that
in $\G' := \G[V']$ every vertex
reaches at least $(\log n)^{(C-3)\gamma}$ vertices during $[0,p/2]$ and
is reached by at least the same number of vertices during $[p/2,p]$.
To this end, we fix a random \emph{balanced} partition $V_1 \cup V_2 \cup \ldots \cup V_C$
of the vertex set $V$ of $\G$ into $C$ parts ('balanced' means that each part has size either $\lceil n/C \rceil$ or $\lfloor n / C \rfloor$); this partition is independent of $\G$. 
We also split the time interval $[0,p]$ into $D := 2C$ 
equal subintervals $T_i$, $i \in [D]$, where $T_i := [(i-1)p/D,\; ip/D]$. 

We will identify subsets $V_i' \subseteq V_i$ such that $|V_i \setminus V_i'| \in o(n)$ 
and for every vertex $v \in V_i'$ we will construct a reachability
tree rooted at $v$ with the desired number of vertices.
Such a tree will have a special structure. Namely,
(i) the $k$-th level of the tree will be entirely in $V_{i+k}'$ 
(for convenience, indices are taken modulo $C$, i.e., $V_{C+1} = V_1$, etc.); and (ii) the time labels of the edges connecting the tree vertices at level $k-1$ with the tree vertices at level $k$ will belong to the $k$-th time interval $T_k$. 
This construction will use the first $C$ time intervals. The last $C$ time intervals will be used in a similar way
to construct reachability trees witnessing that every vertex in $V' := \cup_{i \in [C]} V_i'$ is reachable by the desired number of vertices in $V'$. The argument for the later is the same up to going backward in time, so we provide details only for the first half of the time intervals.

In order to implement the above strategy, we will analyze the structure of the underlying graphs 
of the temporal graphs $\G_{T_i}$, $i \in [D]$ over our fixed balanced vertex partition. Notice that the underlying graph of $\G_{T_i}$ is the graph whose edges are exactly the edges of $\G$ with time labels in $T_i$. Furthermore, each such underlying graph is an \ER{} graph $G_{n,p/D}$. 
The two main technical statements that we need about these underlying graphs and their interaction
with the partition are as follows.

\begin{restatable}{lemma}{multiEpochExpectedNeighbours}
	\label{lm:multi-epoch-expected-neighbours}
	Let $G_1, G_2, \ldots, G_D \sim G_{n, \frac{p}{D}}$ be, not necessarily independent, \ER{}
	graphs on the common vertex set~$V$. Then \aas there exist subsets $V_i' \subseteq V_i$, $i \in [C]$, such that for every $i \in [C]$
	\begin{itemize}
		\item $|V_i\setminus V_i'|\in o(n)$, and
		\item in every $G_j$, $j \in [D]$, every vertex in~$V_i'$ has at least $(\log n)^\gamma$ neighbours in $V'_{i+1}$.
	\end{itemize}
\end{restatable}

\begin{restatable}{lemma}{fewCollisions}
	\label{lm:few-collisions}
	Let $U$ and $W$ be any two fixed parts of the partition $\{ V_i \}_{i \in [C]}$.
	Let $G \sim G_{n, \frac{p}{D}}$ and let $A \subseteq U$ be a set of size 
	at most $(\log n)^{C\gamma}$, chosen independently from the edges of $G$ between $U$ and $W$.
	Then, with probability at least $1-o(1/n)$, in graph $G$ no vertex in $W$ has three neighbors in $A$
	and at most one vertex has two neighbors in $A$.
\end{restatable}

Using these tools, we now prove \cref{lm:bootstrap-reachability}. The proofs of the above two lemmas are provided later on in this subsection.

\bootstrapReachability*
\begin{proof}
	Let $G_i$ be the underlying graph of $\G_{T_i}$, $i \in [D]$. Then all these underlying graphs 
	are distributed according to $G_{n,p/D}$ and, thus, by~\cref{lm:multi-epoch-expected-neighbours} \aas there are subsets $V'_i \subseteq V_i$, $i \in [C]$ satisfying the properties stated in the same lemma.

	Consider some $v\in V'_i$, and let $\ell \in [C]$.
	A temporal path in $\G$ starting at~$v$ is a \emph{clocked path from $v$ of length $\ell$}
	if for every~$j \in [\ell]$ the~$j$-th edge of the path 
	connects a vertex in $V'_{i+j-1}$ with a vertex in $V'_{i+j}$,
	and its time label belongs to $T_j$ (i.e., the edge belongs to $G_j$).
	We will prove that \aas clocked paths from $v$ of length at most $C$ reach at least 
	$(\log n)^{(C-3)\gamma}$ vertices.
	
	Assume this is not the case.
	Then we will show by induction that clocked paths from $v$ of length $j \in [C]$ reach at least $(\log n)^{(j-1)\gamma} +1$ vertices in $V'_{i+j}$, leading to a contradiction as the clocked paths from $v$ of length at most $C-2$ would suffice to reach the desired number of vertices.
	
	The base case where $j=1$ is immediate, since in $G_1$ vertex $v$~has at least $(\log n)^\gamma \geq 2$~neighbors in~$V'_{i+1}$.
	For the induction step, let $A'\subseteq V'_{i+j}$ be the set of vertices reached by~$v$ via clocked paths of length~$j$.
	By the induction hypothesis, $\abs{A'} \geq (\log n)^{(j-1)\gamma}+1$.
	In graph $G_{j+1}$, each of the vertices of~$A'$ has at least $(\log n)^\gamma$~edges to $V'_{i+j+1}$,
	and each of these edges then completes a clocked path from $v$ of length~$j+1$.
	If there were no collisions (i.e., multiple edges leading to the same vertex in~$V'_{i+j+1}$),
	then $v$~would thus reach $(\log n)^{j\gamma}+(\log n)^\gamma$ 
	vertices in~$V'_{i+j+1}$, which is more than enough. It remains to bound the number of collisions.
	
	Ideally, we would like to apply \cref{lm:few-collisions} to~$A'$.
	However, in the proof of \cref{lm:multi-epoch-expected-neighbours}, the set $V'_{i+j}$ (and thus~$A'$) was selected based on the edges of the graphs $G_s$, $s \in [D]$, between $V_{i+j}$ and~$V_{i+j+1}$, thus violating the independence assumption of \cref{lm:few-collisions}.
	To remedy this, let $B' \subseteq V_{i+j-1}'$ contain all vertices reached from~$v$ via clocked paths of length~$j-1$ and let $A \subseteq V_{i+j}$ contain all those vertices that are connected to~$B'$ in~$G_j$.
	Then clearly $A \cap V'_{i+j} = A'$, and the choice of~$A$ is now independent of the edges between $V_{i+j}$ and~$V_{i+j+1}$ in any of the graphs $G_s$, $s \in [D]$.
	By our assumption above, $\abs{B'} < (\log n)^{(C-3)\gamma}$.
	Thus, $\expect[\abs{A}] < (\log n)^{(C-3)\gamma + \alpha} + 1 \leq (\log n)^{(C-1)\gamma}$.
  By a Chernoff bound (Lemma~\ref{lem: chernoff}~\emph{(\ref{chernoff 2})}) and using $(\log n)^{C\gamma} > 6\expect[\abs{A}]$, we have
	\[
	\prob\left[\abs{A} > (\log n)^{C \gamma}\right] < 2^{-(\log n)^{C \gamma}} \in o(1/n).
	\]
	Thus, with probability $1-o(1/n)$, we may apply \cref{lm:few-collisions} to $A$ and $G_{j+1}$ to conclude that in $G_{j+1}$ there is at most one collision between all edges leading from~$A$ to~$V_{i+j+1}$.
	Since $A' \subseteq A$ and $V'_{i+j+1} \subseteq V_{i+j+1}$, the same then holds for the edges from~$A'$ to~$V'_{i+j+1}$.
	In other words, $A'$~has at least $(\log n)^{j\gamma}+(\log n)^\gamma -1 \geq (\log n)^{j\gamma} +1$ neighbors in $V'_{i+j+1}$ in graph $G_{j+1}$, and thus the induction step succeeds with probability $1-o(1/n)$.
	
	By the union bound, this argument holds simultaneously for all~$j \in [C]$ and every possible choice of~$v \in V'$ a.a.s.
	This concludes the proof that \aas in $\GGG[V']$ every vertex reaches at least $(\log n)^{C-3}\gamma$~vertices during the first~$C$~time intervals, i.e., during $[0,p/2]$.
	The second part of the claim is proven analogously, now going backward in time but still in the same direction through the parts of the partition.
\end{proof}

\paragraph*{Properties of \ER{} Graphs over Balanced Partitions.}
\label{sec:ERparatition}

The purpose of what follows is to prove \cref{lm:multi-epoch-expected-neighbours} and \cref{lm:few-collisions}. 
We start with some auxiliary statements.
Recall that $C \geq 3$, $\frac{1}{2}<\gamma<\alpha<1$, $p = 2C^2 \frac{(\log n)^\alpha}{n}$, and $D = 2C$. Let $\beta$ be such that $\gamma<\beta<\alpha$, and
let $U$ and $W$ be two arbitrary but fixed parts of the balanced partition $\{ V_i \}_{i \in [C]}$.

\begin{lemma}
	\label{lm:few-sleepers}
	In $G \sim G_{n, \frac{p}{D}}$ \aas
	at most $n \exp\left(-(\log n)^\alpha/12\right)$ vertices in $U$ have fewer than $(\log n)^\beta$ neighbours in $W$.
\end{lemma}

\begin{proof}
	For any vertex~$u \in U$, its number~$\partial(u)$ of neighbours in~$W$
	is the sum of $|W|$ indicator variables.
	Thus, 
	$\expect[\partial(u)] = |W| \frac{p}{D}\geq
	\left\lfloor\frac{n}{C}\right\rfloor C \frac{(\log n)^\alpha}{n} 
	>\frac{1}{2}(\log n)^\alpha$.
	Using a Chernoff bound (Lemma~\ref{lem: chernoff}~\emph{(\ref{chernoff 3})}),
	\begin{align*}
		\prob[\partial(u) < (\log n)^\beta]
		&< \exp \left( - \frac{\expect[\partial(u)]}{2} \left(1-\frac{(\log n)^\beta}{\expect[\partial(u)]}\right)^2 \right)
		\\ &< \exp \left( - \frac{\expect[\partial(u)]}{3}\right)
		< \exp \left( - \frac{(\log n)^\alpha}{6} \right).
	\end{align*}
	
	As the edges to~$W$ for different $u \in U$ are independent,
	we can further apply a Chernoff bound to estimate the number of vertices with fewer than $(\log n)^\beta$ neighbours.
	For this purpose, let $S=\{u \in U \mid  \partial(u) < (\log n)^\beta\}$.
	Then $\E[\abs{S}]<\left\lceil \frac{n}{C} \right\rceil \cdot \exp \left( - \frac{(\log n)^\alpha}{6}\right)$, and
	by denoting $\delta := \frac{n \exp \left( - \frac{(\log n)^\alpha}{12} \right)}{\E[\abs{S}]}-1 \in \omega(1)$,
	using a Chernoff bound (Lemma~\ref{lem: chernoff}~\emph{(\ref{chernoff 1})}) gives 
	\begin{align*}
		\prob\left[\abs{S}>n\cdot \exp \left( - \frac{(\log n)^\alpha}{12}\right) \right]
		&=
		\prob[\abs{S}>(1+\delta)\E[\abs{S}]]
		\\&<
		\exp\left(-\frac{\delta^2 \E[\abs{S}]}{2+ \delta}\right)
		<
		\exp\left(-\frac{(1+\delta) \E[\abs{S}]}{2}\right)
		\\ &=
		\exp\left(-\frac{n\exp \left(- \frac{(\log n)^\alpha}{12} \right)}{2}\right)
		\in o\left( \exp\left(-\sqrt{n}\right)\right),
	\end{align*}
	which implies the lemma.
\end{proof}

\begin{lemma}
	\label{lm:no-removal-impact}
	Let $G \sim G_{n, \frac{p}{D}}$ and let $S \subseteq W$ be a set of at most
	$Dn \exp\left(-(\log n)^\alpha/12\right)$ vertices
	chosen independently from the edges between $U$ and $W$.
	Then \aas no vertex in $U$ has more than $\frac{1}{2}(\log n)^\beta$ neighbours in~$S$. 
\end{lemma}
\begin{proof}
	Let $u \in U$ and let $\sigma(u)$ denote the number of neighbours of~$u$ in~$S$.
	Observe that $\sigma(u)$~is a sum of $\abs{S}$~independent indicator variables.
	We have that
	$$
	\expect[\sigma(u)] = \frac{p}{D}\abs{S} 
	< 2C^2 (\log n)^\alpha \exp\left(-(\log n)^\alpha/12\right).
	$$
	Let us fix a positive constant $h$ such that 
	$\mu := \expect[\sigma(u)] < \exp\left(-4h(\log n)^\alpha\right)$,
	and let $\delta := \frac{(\log n)^\beta}{2\mu}-1 \in \omega(1)$.
	Then using a Chernoff bound (Lemma~\ref{lem: chernoff}~\emph{(\ref{chernoff 1})}), we derive:
	\begin{align*}
		\prob[\sigma(u) > (\log n)^\beta/2] = 	\prob[\sigma(u) > (1+\delta)\mu]
		&<
		\left( 
		\frac{e^{\delta}}{(1+\delta)^{1+\delta}}
		\right)^\mu
		\\&<
		\left( 
		\frac{e^{1+\delta}}{(1+\delta)^{1+\delta}}
		\right)^\mu
		\\ &=
		\frac{
			\exp\left( (\log n)^\beta/2 \right)
		}{
			\left(\frac{(\log n)^\beta}{2 \mu} \right)^{(\log n)^\beta/2}
		}
		\\ &<
		\frac{
			\exp\left( (\log n)^\beta/2 \right)
		}{
			\left( (\log n)^{\beta}/2 \cdot \exp \left( {4h(\log n)^\alpha}\right)  \right)^{(\log n)^\beta /2}
		}
		\\&<
		\frac{
			\exp\left( (\log n)^\beta/2 \right)
		}{
			\left(  \exp \left( {4h(\log n)^\alpha}\right)  \right)^{(\log n)^\beta /2}
		}
		\\ &=
		\exp\left( \frac{(\log n)^\beta}{2} \right)
		\exp \left( - {2h(\log n)^{\alpha +\beta}} \right)
		\\ &=
		\exp \left(
		\frac{(\log n)^\beta}{2}
		- {2h(\log n)^{\alpha+\beta}}
		\right)
		\\ &<
		\exp \left(- {h(\log n)^{\alpha + \beta}} \right)
		\in o\left(\frac{1}{n}\right).
	\end{align*}
	Applying the union bound over all~$\bigO(n)$~vertices $u \in U$ concludes the proof.
\end{proof}

\begin{corollary}
	\label{lm:no-removal-cascade}
	Let $G \sim G_{n, \frac{p}{D}}$ and let $S \subseteq W$ be a set of at most
	$Dn \exp\left(-(\log n)^\alpha/12\right)$ vertices
	chosen independently from the edges between $U$ and $W$.
	Then \aas every vertex in~$U$ with at least $(\log n)^\beta$ neighbours in~$W$
	has at least $(\log n)^\gamma$ neighbors in $W \setminus S$.
\end{corollary}
\begin{proof}
	A vertex with at least $(\log n)^\beta$ neighbours in~$W$
	but fewer than $(\log n)^\gamma$ neighbours in~$W \setminus S$
	must have more than $(\log n)^\beta - (\log n)^\gamma > \frac{1}{2}(\log n)^\beta$ neighbours in~$S$.
	By~\cref{lm:no-removal-impact},
	there are \aas no such vertices.
\end{proof}

We are now ready to prove the following lemma.

\multiEpochExpectedNeighbours*
\begin{proof}
	For every $i \in [C]$, let $B_i \subseteq V_i$ contains those vertices that have fewer than $(\log n)^\beta$ neighbours in~$V_{i+1}$ in \emph{at least one} of the $D$ graphs $G_1, G_2, \ldots, G_D$. Set $V'_i:=V_i\setminus B_i$.
	
	By~\cref{lm:few-sleepers}, \aas
	each $B_i$ has size at most $D n \exp\left(-(\log n)^\alpha/12\right)  \in o(n)$.
	The choice of $B_{i+1}$ is independent of edges between $V_{i}$ and $V_{i+1}$,
	hence we can apply \cref{lm:no-removal-cascade} and the union bound
	over all $D$ graphs to obtain
	that \aas in each graph every vertex in $V'_i$ has at least
	$(\log n)^\gamma$ neighbours in $V'_{i+1} = V_{i+1} \setminus B_{i+1}$.
\end{proof}

For the proof of \cref{lm:few-collisions}, it will be convenient to have the following claim at hand.

\begin{claim}
	\label{lm:binomial-bound}
	Let $X \sim \operatorname{Bin}(N, q)$ for some~$N \in \NN$ and $2Nq \leq 1$.
	Then 
	\[
	\prob[X \geq k] \leq  2 (Nq)^k .
	\]
\end{claim}
\begin{proof}
	\begin{align*}
		\prob[X \geq k] &= \sum_{i=k}^N \binom{N}{i} q^i (1-q)^{N-i}
		\leq \sum_{i=k}^N \binom{N}{i} q^i
		\leq \sum_{i=k}^N (Nq)^i
		= (Nq)^k \sum_{i=0}^{N-k} (Nq)^i
		\\ &\leq (Nq)^k \frac{1}{1-Nq}
		\leq 2(Nq)^k \,.\qedhere
	\end{align*}
\end{proof}

\fewCollisions*
\begin{proof}
	By \cref{lm:binomial-bound}, the probability that any fixed vertex in $W$ has three or more neighbors in~$A$ is at most
	\[
	2 \left((\log n)^{C \gamma} \frac{p}{D}\right)^3
	=
	2\left(\frac{C (\log n)^{C \gamma + \alpha}}{n}\right)^3
	\in o(1/n^2).
	\]
	Thus, by the union bound, the probability that any vertex in~$W$ has three or more neighbors in $A$ is in $o(1/n)$.
	
	Similarly, the probability that any fixed vertex in~$W$ has two or more neighbors in~$A$ is at most
	\[
	2 \left((\log n)^{C \gamma}\frac{p}{D}\right)^2.
	\]
	Applying \cref{lm:binomial-bound} again gives that the probability of having two or more vertices in~$W$ that each have two or more neighbors in~$A$ is at most
	\begin{align*}
	2\left(|W| \cdot 2\left((\log n)^{C \gamma} \frac{p}{D}\right)^2\right)^2 &\leq
	2\left(2 \ceil*{\frac{n}{C}} \left((\log n)^{C \gamma} \frac{p}{D}\right)^2\right)^2
	\\ &= 8 \ceil*{\frac{n}{C}}^2 \left((\log n)^{C\gamma} \frac{p}{D}\right)^4 
	\in o(1/n)\,. \qedhere
	\end{align*}
\end{proof}

\printbibliography

\end{document}